# Axonal microstructure and compartmentalization impact the orientation and time dependence of mesoscopic transverse relaxation

**Anders Dyhr Sandgaard[1], Rafael Neto Henriques[2], Noam Shemesh[2], Sune Nørhøj Jespersen[1,3]**

[1]Center of Functionally Integrative Neuroscience, Department of Clinical Medicine, Aarhus University, Denmark

[2]Champalimaud Research, Champalimaud Centre for the Unknown, Lisbon, Portugal

[3]Department of Physics and Astronomy, Aarhus University, Denmark

**Corresponding Author**: Anders Dyhr Sandgaard.

**Mail**: anders@cfin.au.dk

# 1| Abstract

In biological tissue, MR transverse relaxation stems from mechanisms spanning multiple scales, from molecular dipole-dipole interactions to mesoscopic field variations driven by tissue microstructure. While mesoscopic relaxation reflects cellular organization, its dynamics in white matter, specifically its dependence on axonal orientation and echo time, remain less investigated. This study employs theoretical frameworks and Monte-Carlo simulations using 3D EM-based white matter (WM) geometries to investigate how compartmentalization and realistic morphology drive these effects. Specifically, we simulate intra-axonal relaxation driven by magnetic fields induced by realistic axonal myelin sheaths and intra-axonal spheres as a model of iron-containing mitochondria. Our results confirm that orientation dependence of mesoscopic relaxation in WM is detectable and agrees with experimental observations. The time-dependence aligns with 1-dimensional short-range structural disorder, but at clinical echo times, this signature may be masked by dominant molecular relaxation. This work moves beyond idealized models to aid the development of more specific biophysical models of mesoscopic relaxation to achieve better biomarkers for neurodegenerative disease.

# 2| Introduction

In biological tissue, the MRI signal is inherently sensitive to water diffusion and microscopic magnetic field variations induced by tissue microstructure. An effective way to probe these variations is by characterizing their impact on transverse relaxation, which provides a window into the underlying tissue architecture. Transverse relaxation describes the loss of phase coherence among spins in the plane perpendicular to the main magnetic field $B_0$ in MRI, and results in decay of the signal over time. It is governed by several biophysical mechanisms, including molecular motion, tissue composition, and susceptibility-induced magnetic field inhomogeneities.

The different mechanisms contributing to relaxation can be broadly categorized into three spatial contributions depending on the scale of the effect. If we describe the normalized MR signal decay $\exp(-\eta(t))$ by a time-dependent transverse relaxation decay function $\eta(t)$, the net decay from the three different length scales can be written as a sum[1]

$$\eta(t) = \eta^{\text{Mol}}(t) + \eta^{\text{Meso}}(t) + \eta^{\text{Macro}}(t).$$

Here $\eta^{\text{Mol}} = R_2^{mol} t$ captures the relaxation on the molecular scale[2], such as the dipole-dipole interaction between the spins of water molecules. Because molecular magnetic field fluctuations occur on time scales much shorter than typical MRI echo times, molecular relaxation can be treated as an effective mono-exponential decay with rate $R_2^{mol}$. On the macroscopic scale[3], $\eta^{\text{Macro}} \sim t^2$describes relaxation arising from static field inhomogeneities, for which diffusion is negligible over the measurement time[1]. Between these two extremes, $\eta^{\mathrm{Meso}}$ describes relaxation on the mesoscopic scale. It depends on how water diffuses across magnetic field gradients generated by heterogeneous tissue environments. The fields may be induced by axons[4,5], iron deposits[6,7], blood vessels[8–13] or other contributors. While molecular and macroscopic relaxation $\eta^{\mathrm{Mol}} + \eta^{\mathrm{Macro}}$ accounts for most of the measured signal relaxation in MRI, $\eta^{\mathrm{Meso}}$ is of particular interest since it provides a window into probing both magnetic and structural features of the biological tissue. However, unlike the other two regimes, its time dependence can be more complex because diffusion samples spatial scales over which the magnetic field varies significantly due to tissue microstructure. For that reason, mesoscopic relaxation has gained a lot of attention throughout the past decades, with efforts focused on understanding it and developing strategies to isolate it from the dominant relaxation mechanisms[14,15].

Two of the most promising ways to isolate mesoscopic relaxation are through its time dependence and its variation with respect to the external magnetic field orientation. We will now give a brief introduction to each of them individually, but for a more in-depth discussion, we refer to the review by Kiselev and Novikov[1].

*Orientation dependence of mesoscopic transverse relaxation*

In brain, variations in net relaxation can be seen, when comparing the signal decay in white matter (WM) across different WM regions[16–19]. This is because WM consists largely of axon bundles, and they exhibit microscopic magnetic anisotropy due to both their orientational coherence and the tensorial susceptibility of myelin lipids[20–28]. This anisotropy leads to orientation-dependent phase shifts and transverse relaxation rates, meaning the MR signal varies with the subject's head orientation in the scanner. Hence, this modulation of the signal relaxation provides a means to probe mesoscopic-scale effects. If, however, these effects are not properly accounted for, they can instead complicate the interpretation of relaxation-based parameters and challenge their use as reliable biomarkers for neurodegenerative disease. In addition, other cellular components such as hemoglobin-rich vasculature[11,12,29–31], iron-containing neuroglia[32] and mitochondria[33], are also likely to contribute to orientation-dependent relaxation due to their spatial organization in and around axons. Recently, molecular dipole–dipole interactions between water and bound protons in the myelin sheath[2,34] have also been proposed as a potential source of orientation dependence (discussed later).

To better understand the origin of this orientation dependence, biophysical models have been developed to describe transverse relaxation analytically[10,12,15,30,31,35–38]. These models often assume low volume fractions of magnetic inclusions and idealized geometries, such as parallel hollow cylinders to mimic axons, or randomly oriented to model microvasculature, and spherical inclusions to mimic iron-rich cells. Recent frameworks have even extended the models to include multiple sources simultaneously, for example by modeling both myelin, vessels and iron inclusions as randomly distributed spheres[39,40].

However, oversimplified assumptions about geometry or diffusion can lead not only to predictions that fail to match experimental observations, but also to models that fit the data well while yielding misleading parameter estimates. For an example of the former, idealized cylindrical models predict that transverse relaxation outside a cylinder varies as $\sin^4(\theta)$, with $\theta$ denoting the angle between the axon axis and the external magnetic field $B_0$, while no transverse relaxation is induced inside the cylinder. In contrast, Figure 1 illustrates how microscopic field perturbations generated by "synthetic axons" with different surface morphologies give rise to orientation-dependent Larmor frequency variance that differs from $\sin^4(\theta)$. For example, when the axon radius varies randomly both radially and axially, the intra-axonal Larmor frequency variance exhibits a dipolar contribution proportional to $(1 - 3\cos^2(\theta))^2$, which also modifies the extra-axonal variance, yielding an orientation dependence of the form $(a \cdot \sin^2(\theta) + b \cdot (1 - 3\cos^2(\theta)))^2$. When used as simplified representations of realistic axonal geometry, these examples show that realistic features - such as variations in cross-sectional area, axial undulations, and surface irregularities - produce orientation-dependent frequency variance and consequently orientation-dependent transverse signal decay. These effects cannot be captured by idealized, perfectly cylindrical geometries (Additional examples and quantitative details are provided in Supplementary Material S1).

Recent work[41] further demonstrates that more realistic WM morphologies generate magnetic field variations inside axons, leading to deviations from the $\sin^4(\theta)$ prediction. Moreover, experimental studies[16,17,19,42–46] consistently suggest the presence of a dipolar contribution, $(1 - 3\cos^2(\theta))^2$, to transverse relaxation - an effect that cannot be explained by models based solely on ideal cylinders, orientation dispersion, or randomly positioned spheres. Although previous studies[42] have attributed this effect to myelin susceptibility anisotropy, this explanation remains unvalidated, since axially symmetric susceptibility aligned with cylindrical layers does not predict such behavior[27].

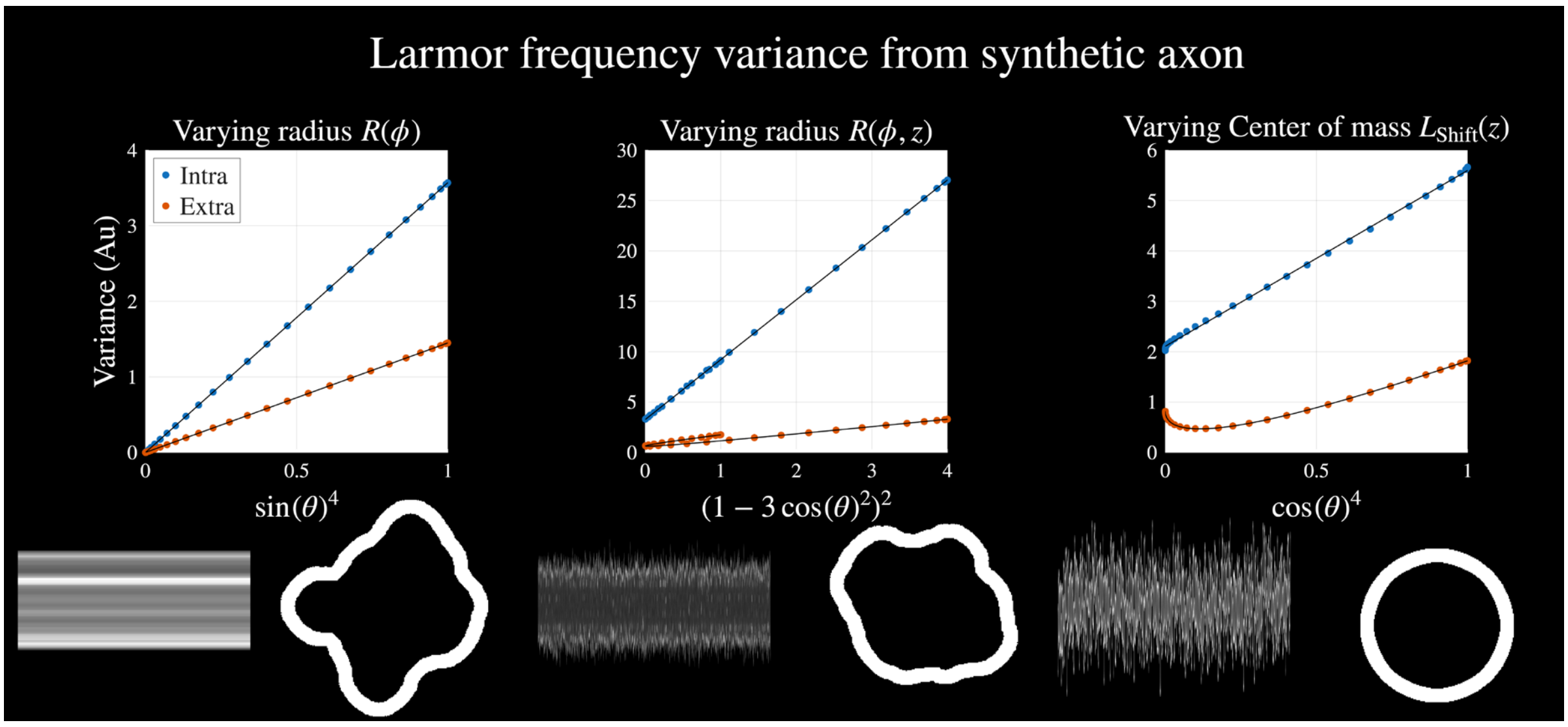

*Figure 1 Intra- and extra-compartmental Larmor Frequency variance are shown for three different perturbations of a normal hollow cylinder. The left plot shows the variance if the radius varies randomly but is fixed along the axis. In the middle plot, the tube radius varies randomly both transverse and axially. The right plot shows the Larmor frequency variance, where each axial slice has a center of mass shifted away from the common center, which demonstrates the effect of axial undulations. See supplementary material S1 for a more detailed analysis of how perturbations of a normal hollow cylinder affect the Larmor frequency variance. The extra-axonal variance appears lower in this visualization because the surrounding extra-cellular space is significantly larger than the axonal compartment.*

*Time dependence of mesoscopic transverse relaxation*

Turning the attention to the time dependence of mesoscopic relaxation: Not only the orientation, but also the time dependence of the signal's mesoscopic relaxation depends in general on the structural disorder of the microstructure[47]. Disorder here refers to deviations from regularity in the spatial arrangement of cellular components. Disorder is typically defined by a characteristic length scale, the correlation length $l$, which is the length beyond which correlations in spatial positions become negligible. This gives rise to an associated correlation time $\tau$ for spins diffusing with diffusivity $D$, corresponding to the time to diffuse past the correlation length $\tau \sim l^2/2D \sim 1-10$ ms, for most biological tissues[48]. Ruh et al. derived[14] that the time-dependent mesoscopic transverse relaxation decay function $\eta^{\mathrm{Meso}}(t)$, has a power law scaling $\eta^{\mathrm{Meso}}(t) \sim (t/\tau)^{-v+2}$ if $v \neq 1$, and $\eta^{\mathrm{Meso}}(t) \sim t/\tau \cdot \ln(t/\tau)$ if $v = 1$. This result applies when the signal decay is well approximated by the second signal cumulant and measured at times $t$ much longer than the microstructure's correlation time $\tau$. Here $v = (p+d)/2$ is a dynamical exponent[47], where $p$ defines the structural disorder class and $d$ the effective dimensionality of the diffusion process, e.g., $d = 3$ for diffusion outside randomly positioned spheres or $d = 1$ inside a long axon. For example[47], short-range disorder refers to systems in which spatial correlations decay sufficiently fast with distance (e.g., exponentially), such that no long-range correlations are present. A short-range media has $p = 0$ and causes a time-dependent signal that decays as $\exp\left[-(t/\tau)^{\frac{4-d}{2}}\right]$. Hyperfluctuating disorder (p = –1), on the other hand, occurs in systems such as disordered arrays of long cylinders where the microstructure becomes increasingly correlated across longer distances[49]. Such disordered media can lead to a time dependent signal $\exp\left[-(t/\tau)^{\frac{5-d}{2}}\right]$ for $d = 1,2$ and $\exp\left[-\left(\frac{t}{\tau}\right)\cdot\ln\left((t/\tau)\right)\right]$ for $d = 3$. It has been shown that intra-axonal space exhibit features of 1D short-range disorder due to variations in its axonal morphology[50], such as fluctuations in diameter and undulations along the axonal axis. These structural irregularities break the idealized cylinder models and thus influence the time dependence of transverse relaxation. However, each axon is also embedded within a dense and disordered environment of neighboring axons, cells and vasculature, and extracellular space, potentially introducing additional sources of short range or hyperfluctuating disorder from the extrinsic magnetic field perturbations they generate. This interplay between internal and external disorder complicates the relaxation dynamics and may give rise to signal decay behaviors that deviate from those predicted by simplified or isolated geometries.

Despite theoretical predictions and simulation-based studies[14,41], such time-dependent effects of transverse relaxation in realistic tissue microstructure remain largely unexplored. Moreover, it is still unclear whether these effects are detectable with practical MRI acquisition protocols. A deeper understanding of how microstructural disorder shapes relaxation dynamics is needed to determine whether such time dependencies can be meaningfully captured and interpreted in vivo.

*Aim of study*

**The aim of this study** is therefore to identify which specific microstructural features are most influential in shaping mesoscopic transverse relaxation, thereby guiding future modeling efforts based on realistic WM architecture. To do so, we investigate the functional dependence of mesoscopic transverse relaxation on both time and orientation, as a means to probe the magnetic microstructure of WM. Using a newly developed Monte Carlo (MC) simulation framework, we model the mesoscopic MR signal for both multi-gradient-echo (MGE) and asymmetric spin-echo (ASE) sequences in mesoscopically sized WM segments containing coherently oriented axons and randomly positioned intra-axonal spheres.

**Objective 1)** We derive how mesoscopic transverse relaxation in an axially symmetric microstructure varies with the orientation relative to the main field $\mathbf{B_0}$. We also derive the effect of compartmentalization, as intra-axonal volume consists of many axons with non-exchanging water compartments.

**Objective 2)** We validate our theoretical findings by comparing the simulated intra-axonal mesoscopic relaxation in WM against our theoretical predictions. This is investigated using magnetic field perturbations generated by realistic axonal microstructure, comprising uniformly magnetized, myelinated axons, and spherical inclusions which mimic intra-axonal iron-filled mitochondria[33].

**Objective 3)** We investigate if the simulated intra-axonal mesoscopic relaxation's time dependence exhibits features consistent with 1D short-range structural disorder. This is to see if the mesoscopic relaxation exhibits a unique time-dependence which, in addition to its orientation dependence, can help differentiating it from the pure molecular and macroscopic relaxation.

**Objective 4)** As our findings give insights into the orientation dependent effect of mesoscopic relaxation, we compare it against experimental measurements of the net transverse relaxation to see if it matches reality.

## 3| Theory

The primary objective in the theory section is to investigate how microstructural features can modulate mesoscopic transverse relaxation in the MR signal of biological tissue. Specifically, we examine the mesoscopic relaxation's time-dependence, its relationship to compartmentalization, and its orientation-dependence relative to the main magnetic field $B_0$. While the focus of this study is WM, the theoretical results are not limited to this tissue type and may apply to other biological structures with similar properties.

### *System of consideration*

We consider a magnetized random medium within a mesoscopic volume containing non-exchanging water compartments labelled by $c$. In WM, the many compartments correspond to the hundreds of thousands of individual axons in a voxel[51] that act as discrete, non-exchanging environments for water molecules[52].

The magnetized medium is described by a scalar magnetic susceptibility $\chi(\boldsymbol{r})$ ($|\chi| \ll 1$, and is given relative to the susceptibility of water), and the induced microscopically varying magnetic field as

$$\Delta\mathbf{B}(\boldsymbol{r}) \simeq B_0 \int_{\mathcal{M}} \mathrm{d}\boldsymbol{r}' \, \boldsymbol{\Upsilon}(\boldsymbol{r}-\boldsymbol{r}')\chi(\boldsymbol{r}')\widehat{\mathbf{B}}. \tag{1}$$

Here $\boldsymbol{\Upsilon}(\boldsymbol{r})$ denotes the dipole field tensor and $\mathbf{B}_0 = B_0\widehat{\mathbf{B}}$ is the external field. The induced field gives rise to a microscopically varying Larmor frequency distribution $f(\boldsymbol{r}) \simeq \gamma\widehat{\mathbf{B}}^{\mathbf{T}}\Delta\mathbf{B}(\boldsymbol{r})/2\pi$. For later convenience, we denote the *intra-compartmental* mean frequency, variance and standard deviation as $f_c$, $\sigma_c^2$ and $\sigma_c$ respectively. When the magnetic susceptibility is constant within the magnetized medium, we can rewrite it in terms of a structural indicator function $v(\boldsymbol{r})$, such that $\chi(\boldsymbol{r}) = \chi v(\boldsymbol{r})$, and by a characteristic intrinsic frequency defined as $f_0 = \gamma B_0 \chi / 2\pi$. The characteristic frequency $f_0$, is introduced to highlight the inherent ambiguity between the effects of magnetic susceptibility $\chi$ and the static magnetic field $B_0$ on the MR signal. By combining these parameters, we acknowledge that the microscopic field perturbations $\Delta\mathbf{B}$, are driven by the product of the two, making their effect on the signal decay indistinguishable to either parameter.

The complex MR signal of the water is $S = \langle e^{-i\varphi(t)} \rangle$, where brackets $\langle \ldots \rangle$ signify an average over spins. The phase $\varphi(t) = 2\pi \int_0^t dt' \, \xi(t') f(\boldsymbol{r}_{t'})$ describes the time along its stochastic trajectory $\mathbf{r}_{t'}$. The pulse sequence is described by a general spin-flip function $\xi(t)$. A Spin-Echo (SE) pulse sequence is defined by a change in sign at half the echo time $T_E$, while a Free-Induction-Decay (FID) is always unity. When considering water in a single compartment, we can write $S_c(t) = \langle e^{-i\varphi(t)} \rangle_c$, which then only averages across water in compartment $c$. For simplicity, we focus on the mesoscopic relaxation effect $\eta^{Meso}(t)$ from $\Delta\mathbf{B}$, and assume that molecular and macroscopic relaxation $\eta^{Mol}(t) + \eta^{Macro}(t)$ is uniform across all compartments (see Supplementary Material S3 for cases where this does not hold). Consequently, we neglect molecular and macroscopic relaxation throughout our initial derivations to focus exclusively on mesoscopic effects, re-instating them only when necessary.

*Time-dependent mesoscopic relaxation*

We start by addressing the time dependent phase $\varphi_c$ and mesoscopic transverse relaxation $\eta_c$ of water compartment $c$, which could mimic a single intra-axonal compartment in WM. We omit explicitly indicating the time dependence $t$ for notational simplicity.

Because the susceptibility differences in WM are relatively small[1], the signal should in most case be well described by the weak dephasing regime $\sigma_c \tau \ll 2\pi$.[1] When dephasing is weak, we may use perturbation theory to characterize the main effect of $\Delta\mathbf{B}$ on the MR signal through the cumulant expansion[53] (in dMRI, a similar cumulant approach is used to characterize the signal attenuation resulting from a diffusion gradient[54,55]). In particular, we can perform a Taylor expansion on the logarithm of the complex signal $\ln(S)$ in powers of $\varphi_c = \langle \varphi \rangle_c = 2\pi t f_c$ to yield cumulants[53] - quantities that are often better conditioned than the coefficients of a direct Taylor expansion of $S_c$. Physically, the improved conditioning follows because the cumulants here represent statistical moments related to the distribution of phase shifts $\varphi$ in compartment $c$ rather than the explicit moments of the signal $S_c$ itself. The compartmental signal can in general be written as a normalized exponentially decaying function $S_c = \exp(-\eta_c - i\varphi_c)$. In the weak dephasing limit ($\sigma_c \tau \ll 2\pi$), the phase can thus be characterized by the first cumulant $\varphi_c = \langle\varphi\rangle_c = 2\pi t f_c$ for an FID signal[23], and $\varphi_c = 0$ for an SE signal. The first cumulant, $\langle\varphi\rangle_c$ describing the FID signal phase, depends directly on the average induced magnetic field $\Delta\boldsymbol{B}$ in compartment $c$, cf. Eq. (1), and thus scale linearly in time $t$. The mesoscopic transverse relaxation is characterized by the second cumulant $\eta_c = \langle\varphi^2\rangle_c/2$. The second cumulant will be different depending on the pulse sequence $\xi$, and scales with the phase variance $\sigma_c^2$. It is important here to note that while writing the signal in terms of $\varphi_c$ and $\eta_c$ is general, their functional dependence on $\Delta\mathbf{B}$ may change beyond the weak dephasing limit depending on the microstructure and characteristic frequency $f_0$.[1] For example, as the variance $\sigma_c^2$ increases, higher-order terms of the cumulant expansion may become necessary to describe $\varphi_c$ and $\eta_c$. In extreme cases, such as tissues with high iron load or regions adjacent to large vessels, the field perturbations become so strong that the cumulant expansion may fail to converge all together. In these regimes, non-perturbative descriptions are required to accurately model the signal decay.[12,38,56]

In this study of WM, we assume to operate within the weak perturbation limit ($\sigma_c \tau \ll 2\pi$), where the second cumulant provides a robust first-principles description of the mesoscopic relaxation. This means that the second order cumulant $\langle\varphi^2\rangle_c$ is assumed to be sufficient in describing the time dependence of $\eta_c$. As discussed in the introduction, the mesoscopic signal decay $\eta_c$ is generally not mono exponential in time. Instead, the signal decay $\eta_c$ in a water compartment is determined by the interplay between the medium's microstructure, its magnetization (which combined describes the susceptibility-induced field perturbations $\Delta\mathbf{B}$), and the diffusion time.[1] As derived by Ruh et al.[14]: When the decay is weak, $\eta_c \ll 1$, and the effect of diffusion is negligible, i.e. $t \ll \tau$, the signal decay becomes $\eta_c \simeq \sigma_c^2 t^2/2$. When $t \gg \tau$, the functional form of the time dependence[14] $\eta_c \propto t^{-\nu+2}$ $(\nu \neq 1)$ is determined by the type of structural disorder via the dynamical exponent $\nu = (p+d)/2$.[47] Previous studies of the time dependent diffusivity for intra-axonal water have found evidence of 1D short-range disorder due to varying morphology of the axonal sheath along the axon.[50] This would lead to an exponent of $\nu = 1/2$ and a compartmental decay $\eta_c \propto t^{-3/2}$. Having established the expected time dependence of a single compartmental signal $S_c$, the next thing to consider is the effect of measuring the total signal from many compartments at once.

*MR signal of multiple water compartments*

While the previous section provides insights into the mesoscopic decay of a single compartment, accounting for the decay of hundreds of thousands of intra-axonal compartments individually would be a daunting task. We therefore aim to identify the main influence of compartmentalization on the total signal decay across multiple compartments, and here we focus on the FID and SE signal. For generality, we write the total mesoscopic complex FID signal, as a normalized exponentially decaying function $S^{FID} = \exp(-\eta^{FID} - i\varphi^{FID})$, where $\eta^{FID}$ describes the dimensionless time-dependent accumulated mesocopic transverse relaxation of the total signal across all water compartments ($\eta^{FID} + \eta^{Mol} + \eta^{Macro}$ when including the other relaxation terms), and $\varphi^{FID}$ the net accumulated phase at time $t$. The total mesoscopic SE signal would likewise be $S^{SE} = \exp(-\eta^{SE})$. At each time point, we can define an accumulated relaxation rate as $R_2^{FID} = \eta^{FID}/t$ or $R_2^{SE} = \eta^{SE}/t$, respectively.

Focusing on the total FID signal: If we assume the total signal is a sum of compartment signals $S_c$, then the net normalized mesoscopic FID signal can be written as

$$S^{FID} = \exp(-\eta^{FID} - i\varphi^{FID}) = \sum_c w_c S_c = \sum_c w_c \exp(-\eta_c - i\varphi_c), \qquad (2)$$

where the $w_c$ define the signal fractions with $\sum_c w_c = 1$. For later convenience, we introduce the mean *intercompartmental* decay $\overline{\eta} = \sum_c w_c \eta_c$ and phase $\overline{\varphi} = \sum_c w_c \varphi_c$, where $\overline{(\ldots)} = \sum_c w_c (\ldots)_c$ denotes the *intercompartmental* mean. The challenge now is to understand how the net FID decay $\eta^{FID}$ depends on $w_c$, $\eta_c$ and $\varphi_c$.

Analogous to the previous section on time dependence, where we considered the signal decay within a single compartment, we can also use perturbation theory to characterize the main effect of compartmentalization on the FID signal through a cumulant expansion. In particular, we perform a Taylor expansion on the logarithm of the complex signal $\ln(S^{FID})$ in Eq. (2), but now in powers of $\overline{\eta} + i\overline{\varphi}$. Here, the cumulants represent the statistical moments related to the distribution of compartmental offsets from the mean, i.e. $\eta_c - \overline{\eta} + i(\varphi_c - \overline{\varphi})$. Here we consider the cumulant expansion up to second order, to characterize the major effect of compartmentalization. We believe this is also a reasonable assumption for bundles of coherently oriented axons, where the main difference in $\eta_c + i\varphi_c$ across different compartments will come from structural variations in the axonal sheaths[57,58] (in supplementary material S3, we extend our result to include multiple axonal bundles with arbitrary orientations). For convenience, we also define $\overline{\varphi^2} - \overline{\varphi}^2 = \sum_c w_c (\varphi_c - \overline{\varphi})^2$ as the intercompartmental variance of intra-compartmental mean phase shifts, $\overline{\eta^2} - \overline{\eta}^2 = \sum_c w_c (\eta_c - \overline{\eta})^2$ as the intercompartmental variance of intra-compartmental signal decay functions, and lastly the intercompartmental covariance as $\overline{\eta\varphi} - \overline{\eta} \cdot \overline{\varphi} = \sum_c w_c (\eta_c - \overline{\eta})(\varphi_c - \overline{\varphi})$

Rewriting the logarithm of the signal of Eq. (2) with respect to $\overline{\eta} + i\overline{\varphi}$ we get

$$\ln(S^{FID}) = \ln\left( \exp(-\overline{\eta} - i\overline{\varphi}) \sum_c w_c \exp\big(-(\eta_c - \overline{\eta}) - i(\varphi_c - \overline{\varphi})\big) \right).$$

Using the logarithmic rules $\ln(\exp(A) \cdot B) = \mathrm{A} + \ln(B)$, along with the Taylor expansions $\ln(1-x) \approx -x + x^2/2 + \cdots$ (for $x \to 0$) and $\exp(-x) \approx 1 - x + \frac{x^2}{2} + \cdots$ (for $x \to 0$), we find that the logarithm of the signal becomes (to second order)

$$\ln(S^{FID}) \approx \left((-\overline{\eta} - i\overline{\varphi}) - 1 + \sum_c w_c \left(1 - (\eta_c - \overline{\eta}) - i(\varphi_c - \overline{\varphi}) + \frac{1}{2}\left((\eta_c - \overline{\eta}) + i(\varphi_c - \overline{\varphi})\right)^2\right)\right).$$

As $\sum_c w_c = 1$, it cancels out the -1. Second, the sums $\sum_c w_c(\eta_c - \overline{\eta}) = \overline{\eta} - \overline{\eta} = 0$, and $\sum_c w_c(\varphi_c - \overline{\varphi}) = 0$. Writing out the second order term explicitly, we arrive at the four main contributions

$$\ln(S^{FID}) \approx \left((-\overline{\eta} - i\overline{\varphi}) - \frac{1}{2}\sum_c w_c(\varphi_c - \overline{\varphi})^2 + \frac{1}{2}\sum_c w_c\,(\eta_c - \overline{\eta})^2 + i\sum_c w_c\,(\eta_c - \overline{\eta})(\varphi_c - \overline{\varphi})\right).$$

Taking the exponential on each side, we thus get

$$S^{FID} \approx \exp\left(-\overline{\eta} - i\overline{\varphi} - \frac{1}{2}\left(\overline{\varphi^2} - \overline{\varphi}^2\right) + \frac{1}{2}\left(\overline{\eta^2} - \overline{\eta}^2\right) + i(\overline{\eta\varphi} - \overline{\varphi} \cdot \overline{\eta})\right). \tag{3}$$

Here, the variance $\overline{\eta^2} - \overline{\eta}^2$ in the compartmental signal decay functions $\eta_c$ reduces the overall signal decay rate in a manner analogous to the effect of kurtosis[54] in DKI. The covariance term $(\overline{\eta\varphi} - \overline{\varphi} \cdot \overline{\eta})$ also reduces the overall signal phase. However, the two terms scale with the characteristic frequency as $f_0^4$ and $f_0^3$, respectively. Because we assume to be in the weak dephasing regime $\sigma_c \tau \ll 1$, for all compartments, we can neglect $\overline{\eta^2} - \overline{\eta}^2$ and $(\overline{\eta\varphi} - \overline{\varphi} \cdot \overline{\eta})$ is they should have little impact on $\eta^{FID} + i\varphi^{FID}$. Since $\varphi_c = 2\pi f_c t$, i.e. that the compartmental signal phase is described by the spatially averaged Larmor frequency shift $f_c$, cf. Eq. (1), we can rewrite the phase contributions in terms of the Larmor frequency shift $\overline{\varphi} = 2\pi t \overline{f}$ and $\overline{\varphi^2} - \overline{\varphi}^2 = (2\pi t)^2 \left(\overline{f^2} - \overline{f}^2\right)$. This means that intercompartmental Larmor frequency variance $\overline{\varphi^2} - \overline{\varphi}^2$ gives rise to a signal exponentially decaying with $t^2$. The net mesoscopic FID signal decay $\eta^{FID}$ thus becomes

$$\eta^{FID} = \overline{\eta}^{\mathrm{FID}} + \frac{1}{2}\left(\overline{\varphi^2} - \overline{\varphi}^2\right) \tag{4}$$

(FID signal decay)

The mesoscopic decay $\eta^{FID}$ scale quadratically with the characteristic frequency, $f_0 = \gamma B_0 \chi / 2\pi$. This is because $\overline{\eta}$ is just the average of $\eta_c$, which scales with the frequency variance $\sigma_c^2$, while the phase variance, $\varphi_c$ scales linear with $f_0$ since it relates to the mean frequency shift $f_c$. This effectively means that, as the magnetic field strength $B_0$ or intrinsic susceptibility $\chi$ increase, the total signal decay increase quadratically, in agreement with previous experimental results for transverse relaxation in WM[59].

For a mesoscopic SE signal, where the inter-compartmental phase variance will be refocused, the decay $\eta^{SE}(t)$ becomes

$$\eta^{SE} = \overline{\eta}^{SE}. \quad (5)$$

(SE signal decay)

We emphasize that the mean decay $\overline{\eta}$ is different between FID and SE, and should be smaller for SE compared to FID[1], i.e. $\overline{\eta}^{SE} \leq \overline{\eta}^{FID}$, since the intra-compartmental decay $\eta_c$ are partly refocused by the 180-degree RF pulse.

Having established the effect of compartmentalization, we can now tie Eqs. (4) and (5) together with the previous section on the time dependence of mesoscopic relaxation in a single compartment. For long echo times $t \gg \tau$ compared to the correlation time $\tau$, these different terms of Eqs. (4) and (5) scale with time as $\overline{\eta} \propto t^{-2v+2}$ and $\overline{\varphi^2} - \overline{\varphi}^2 \propto t^2$, while for short times where $t \ll \tau, \overline{\eta} \propto t^2$ and $\overline{\varphi^2} - \overline{\varphi}^2 \propto t^2$. Hence, two potential contributions from the internal field with unique time dependences may contribute to the signal's mesoscopic relaxation.

For completeness, adding molecular and macroscopic relaxation gives $\eta^{FID} = \overline{\eta}^{\mathrm{FID}} + \frac{1}{2}\left(\overline{\varphi^2} - \overline{\varphi}^2\right) + \eta^{Mol} + \eta^{Macro}$, while $\eta^{SE} = \overline{\eta}^{SE} + \eta^{Mol}$, since macroscopic relaxation will be refocused in a spin echo pulse sequence. As commented previously by Kiselev and Novikov[1], an important thing to note here is that, in common notation[60], $R_2'$ is defined as the difference between the total transverse relaxation rate measured via the free-induction decay ($\eta^{FID} = R_2^\star t$) and the irreversible relaxation rate measured via a spin-echo ($\eta^{SE}$=$R_2 t$). While $R_2'$ is frequently used as a proxy for reversible relaxation caused by mesoscopic field inhomogeneities[60], its value is not intrinsic to the FID signal alone. Instead, it reflects the signal recovered by the spin-echo refocusing pulse. Thus, $R_2'$ represents the sensitivity difference of these two pulse sequences, captured by the difference in their respective second cumulants, to the underlying microstructural magnetic field variations.

*Orientation dependence of mesoscopic relaxation*

Beyond time dependence, the mesoscopic signal decay may also be isolated through its orientation dependence. As described in the introduction, when the microstructure is anisotropic, such as in WM where axons are symmetrically oriented around a main fiber axis[51], the Larmor frequency shifts $f(\boldsymbol{r})$ vary according to the axons' orientation relative to the external field (see supplementary material S3 for non-axially symmetric distributions). Consequently, the total mesoscopic decay $\eta$ for a general pulse sequence becomes orientation dependent. Here, we demonstrate that for weakly magnetized tissue with axially symmetric microstructure, the orientation dependence of the mesoscopic transverse relaxation decay includes only even-order cosines up to fourth order $(1, \cos^2(\theta), \cos^4(\theta))$. These results are valid for general pulse sequences provided the total relaxation is fully defined by the second cumulant $\eta = \langle \varphi^2 \rangle /2$, and induced by the microstructural magnetic field $\Delta\mathbf{B}$. This also means that the framework consistently describes the net decay even in the presence of compartmentalization, as expressed in Eqs. (4) and (5) (which we discuss later in the section).

The second cumulant of a signal from a sequence described by a general spin flip function $\xi(t)$ (for example a 180-degree RF pulse for a spin echo sequence) is

$$\langle \varphi^2 \rangle = (2\pi)^2 \int_0^t dt \int_0^t dt' \, \xi(t)\xi(t') \langle f(\boldsymbol{r}_t) f(\boldsymbol{r}_{t'}) \rangle \,, \tag{6}$$

$$\simeq (\gamma B_0)^2 \int_0^t dt \int_0^t dt' \xi(t)\xi(t') \int_{\mathcal{M}} d\boldsymbol{r} \int_{\mathcal{M}} d\boldsymbol{r}' \, \widehat{\mathbf{B}}^{\mathbf{T}} \boldsymbol{\Upsilon}(\boldsymbol{r}) \widehat{\mathbf{B}} \widehat{\mathbf{B}}^{\mathbf{T}} \boldsymbol{\Upsilon}(\boldsymbol{r}') \widehat{\mathbf{B}} \langle v^w(\boldsymbol{r}_t) \chi(\boldsymbol{r}_t + \boldsymbol{r}) v^w(\boldsymbol{r}_{t'}) \chi(\boldsymbol{r}_{t'} + \boldsymbol{r}') \rangle \,.$$

Here the indicator function $v^w(\boldsymbol{r})$ specifies the MR visible medium, while $\chi(\boldsymbol{r})$ specifies the location of magnetized microstructure. We made a change of variables such that the dipole kernel does not depend on the time-dependent particle position. We now write out the dipole fields to yield

$$\widehat{\mathbf{B}}^{\mathbf{T}} \boldsymbol{\Upsilon}(\boldsymbol{r}) \widehat{\mathbf{B}} \widehat{\mathbf{B}}^{\mathbf{T}} \boldsymbol{\Upsilon}(\boldsymbol{r}') \widehat{\mathbf{B}} \propto \frac{1}{r^3} \frac{1}{r'^3} \left(1 - 3\left(\hat{\boldsymbol{r}} \cdot \widehat{\mathbf{B}}\right)^2\right) \left(1 - 3\left(\hat{\boldsymbol{r}}' \cdot \widehat{\mathbf{B}}\right)^2\right).$$

For simplicity and without loss of generality in axially symmetric microstructure, we choose the symmetry axis to be $\hat{z}$ and take $\widehat{\mathbf{B}} = [\sin(\theta) \; 0 \cos(\theta)]^T$, so the angular part can be rewritten as

$$\begin{aligned} \left(1 - 3\left(\hat{\boldsymbol{r}} \cdot \widehat{\mathbf{B}}\right)^2\right) \left(1 - 3\left(\hat{\boldsymbol{r}}' \cdot \widehat{\mathbf{B}}\right)^2\right) \\ = 1 \\ - 3\big( (\hat{r}_x'^2 + \hat{r}_x^2) \sin^2(\theta) - (\hat{r}_z'^2 + \hat{r}_z^2) \cos^2(\theta) \\ - 2(\hat{r}_x' \hat{r}_z' + \hat{r}_x \hat{r}_z) \sin(\theta) \cos(\theta) \big) \\ + 9(\hat{r}_x'^2 \hat{r}_x^2 \sin^4(\theta) + \hat{r}_z'^2 \hat{r}_z^2 \cos^4(\theta) \\ + 4(\hat{r}_x'^2 \hat{r}_z^2 + \hat{r}_x^2 \hat{r}_z'^2 + \hat{r}_x' \hat{r}_z' \hat{r}_x' \hat{r}_z') \cos^2(\theta) \sin^2(\theta) \\ + 2(\hat{r}_x'^2 \hat{r}_x \hat{r}_z + \hat{r}_x^2 \hat{r}_x' \hat{r}_z') \cos(\theta) \sin^3(\theta) \\ + 2(\hat{r}_z'^2 \hat{r}_x \hat{r}_z + \hat{r}_z^2 \hat{r}_x' \hat{r}_z') \cos^3(\theta) \sin(\theta)). \end{aligned} \tag{7}$$

Consider the terms above with odd order combinations of sines and cosines. For an axially symmetric microstructure, the second cumulant must satisfy antipodal symmetry (inversion through the origin). While a true antipodal transformation implies $\phi' = \pi - \phi$ and $\theta' = \pi - \theta$, the rotational symmetry of the system makes the azimuthal shift ($\phi$) irrelevant. Consequently, this combination of symmetries requires the distribution to be invariant under the transformation (effectively mirror symmetry across the $xy$-plane). All odd angular terms in the expansion violate this symmetry and therefore cannot contribute All odd angular terms violate this symmetry and therefore cannot contribute. Consequently, the orientation dependence is fully determined by even-order cosine terms only.

$$\eta = a \cdot \cos^4(\theta) - b \cdot cos^2(\theta) + c. \quad (8)$$

This result is valid for a general pulse sequence and should thus describe the mesoscopic relaxation $\eta$ for both $\eta^{FID}$ and $\eta^{SE}$. As the result is independent of diffusion dynamics and spin flip function $\xi(t)$, the functional form is also valid for the Larmor frequency variance (this is demonstrated in Supplementary material S2). For completeness, adding molecular and macroscopic relaxation $\eta^{mol} + \eta^{macro}$ will be absorbed into the parameter $c$. While the total MRI relaxation rates may be high due to these effects, as reflected by the offset $c$, the orientation dependence captured through $a$ and $b$ provides the essential window into the mesoscopic scale.

The amplitudes, except $a$, may be negative as long as $\eta$ is positive overall for all $\theta$. For example, the variance outside a straight cylinder scales as $\sin^4(\theta) = 1 - 2\cos^2(\theta) + \cos^4(\theta)$. A similar orientation dependence occurs inside and outside a long tube whose cross-section varies randomly but does not change along its axis. The variance outside anisotropically positioned spheres (an analogy could be a string of beads[61]) scales as $(1 - 3\cos^2(\theta))^2 = 1 - 6\cos^2(\theta) + 9\cos^4(\theta)$. Hence, different structural features leads to different weights $a, b, c$, and each may in general reflect combined effects of multiple features. Notice all the exemplified structures mentioned here produced a negative amplitude for $\cos^2(\theta)$. While such orientation dependence of Eq. (8) has been proposed before, e.g. by imposing susceptibility anisotropy[42] or dipole-dipole interactions with the myelin sheath[34], the novelty here is that we arrived at it assuming only an axially symmetric and translation invariant microstructure with uniform (scalar) susceptibility.

*In total*

Figure 2 provides an overview of mesoscopic transverse relaxation in realistic WM microstructure from different perspectives. When comparing the description of orientation dependence in Eq. (8) with the compartment- and time-dependence characterized in Eqs. (4) and (5), it becomes evident that while both frameworks are physically valid, they describe different dimensions of the same mesoscopic signal decay process $\eta$. This means that the amplitudes $a, b, c$ in Eq. (8) are inherently time-dependent, containing distinct contributions from both the mean signal decay $\overline{\eta}$, and the phase variance $\overline{\varphi^2} - \overline{\varphi}^2$. Conversely, the terms in Eqs. (4) and (5) must also be orientation-dependent, such that both the mean and variance terms are orientation dependent as in Eq. (8), when the overall microstructure is axially symmetric, but scaled by their own unique amplitudes $a, b, c$.

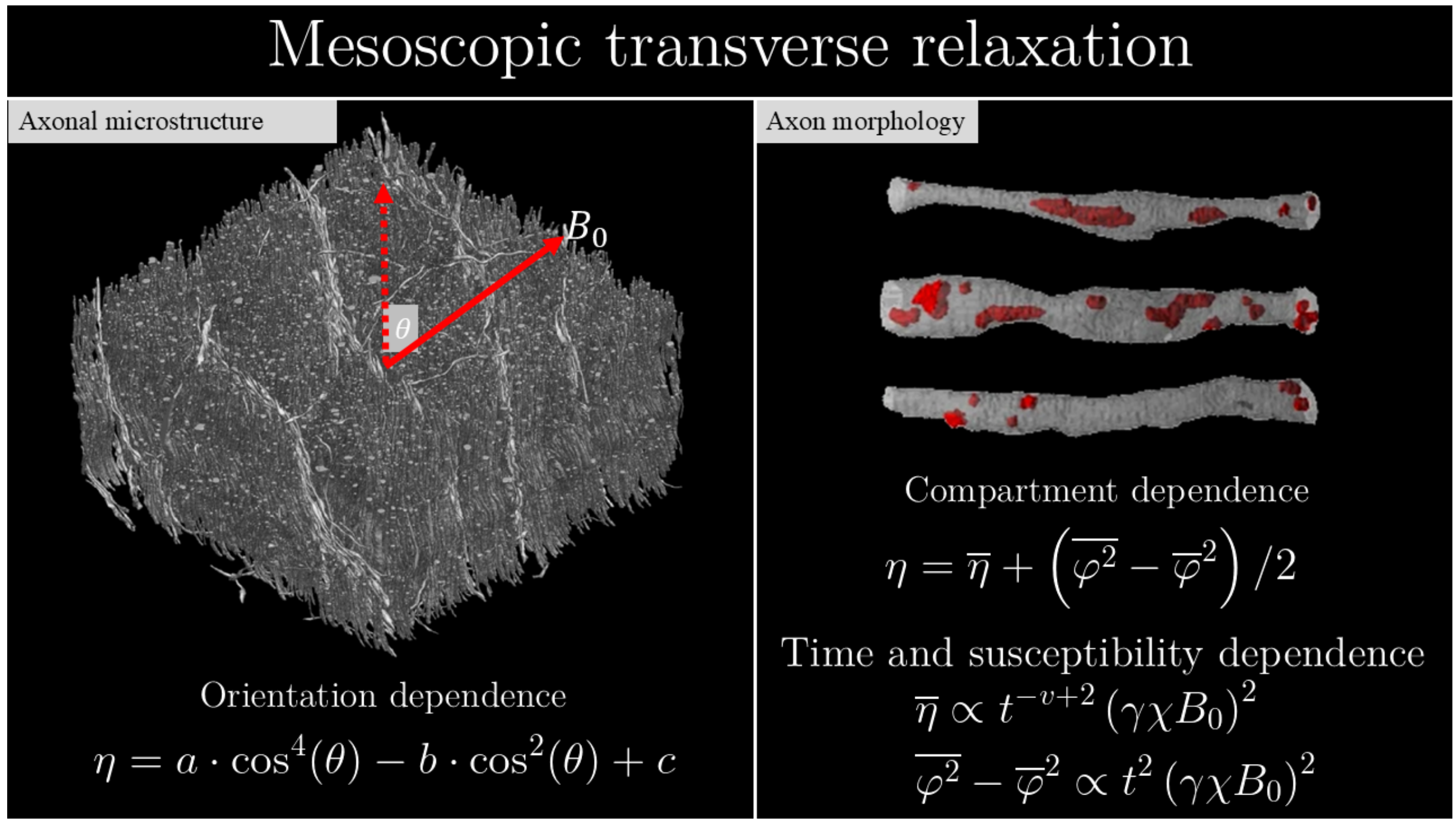

*Figure 2 - Overview of mesoscopic transverse relaxation in realistic WM microstructure. The left panel illustrates the orientation dependence of* $\eta$ *for an axially symmetric microstructure, represented by realistic WM axons. The dotted red arrow represents the symmetry-axis, while the solid red arrow represents the direction of the external field* $B_0$*. The right panel highlights the effect of microstructural compartmentalization and its associated time and susceptibility dependence. Representative axons are shown containing iron-rich mitochondria (red), which both contribute to the local field perturbations. The segmented microstructure to the left is described by Abodollahzadeh et al.*[62] *while the right axons are adapted from Lee et al.*[50]

## 4| Methods

All simulations and analyses were done in MATLAB (The MathWorks, Natick, MA, USA). All animal experiments were preapproved by the competent institutional and national authorities and carried out according to European Directive 2010/63.

Supplementary material contains additional simulations where we computed the Larmor frequency variance from both the realistic and synthetically generated axons created by perturbing the surface of an ideal cylinder. Those simulations allowed us to validate our simulation framework and to investigate how different microstructural features impacted the orientation dependence. Investigating the induced field variance from both individual axons and the whole microstructure also enabled us to assess the importance of magnetic field variances induced from neighboring axons and from different types of magnetic inclusions.

**Simulations: Investigating transverse relaxation inside realistic axonal microstructure**

We designed a set of simulations to examine the orientation dependence of Eqs. (4) and (8) from full WM axonal microstructures of rats from EM containing thousands of axons. Figure 3 gives an overview of the EM substrates used. TBI denotes rats were exposed to a Traumatic Brain Injury, while sham were control rats treated similar to the TBI rats exceptfor the TBI injury. The TBI axons have been shown to exhibit increased beading[63] compared to sham. We considered the individual contribution from magnetic fields induced *i*) by the axons and *ii*) manually introduced intra-axonal spherical inclusions modelling iron-filled mitochondria[33]. Every inclusion, also myelin, was assumed to have scalar susceptibility[20,64] in every simulation. For each substrate, we identified the principal fiber direction and selected a major axon bundle consisting of approximately 2,000 axons aligned predominantly along this direction. As described in our previous study[21], the axonal microstructure in each substrate was defined by an indicator function $v(\boldsymbol{r})$ on a 3D grid with resolution of 0.1 μm³ (see ref [21] for more details).

Due to the inherent ambiguity in decoupling the effects of magnetic susceptibility $\chi$ and the static magnetic field $B_0$ as the measured phase shift $\varphi$ is proportional to the product, we define our parameters in terms of the intrinsic characteristic frequency $f_0 = \gamma B_0 \chi / 2\pi$. This approach ensures considering a realistic range of susceptibilities across both clinical and preclinical field strengths without requiring the isolation of individual physical constants. For the myelinated axons in *i)*, we considered three intrinsic characteristic frequencies approximately $f_m = -[28,\ 65,\ 150]$ Hz.[20,64,65] The local Larmor frequency shift $f(\boldsymbol{r})$ was calculated numerically[66] using Eq. (1), with $f_m(\boldsymbol{r}) = v_m(\boldsymbol{r}) f_m$ replacing $\chi(\boldsymbol{r})$ in the equation. $f(\boldsymbol{r})$ was calculated for 50 unique orientations of $\widehat{\mathbf{B}}$ generated using electrostatic repulsion[67]. As our simulation resolution (0.1 μm³) was too coarse to model effects from small inclusions like ferritin molecules with a radius around 4 nm,[68] we instead modelled iron-containing mitochondria in *ii)* with a radius of approximately 0.5 μm. We chose a volume fraction of 5%, and three characteristic frequencies approximately $f_s = [140,\ 328,\ 768]$ Hz.[7,69,70] Non-overlapping intra-axonal spheres were randomly packed using a previously designed packing generator[23]. We assumed the spherical cellular inclusions were impenetrable to water and their water signals fully relaxed, like myelin water. Again, we calculated the local Larmor frequency shift $f(\boldsymbol{r})$ numerically[66] using Eq. (1), with $f_s(\boldsymbol{r}) = v_s(\boldsymbol{r}) f_s$ for the same 50 orientations of $\widehat{\mathbf{B}}$.

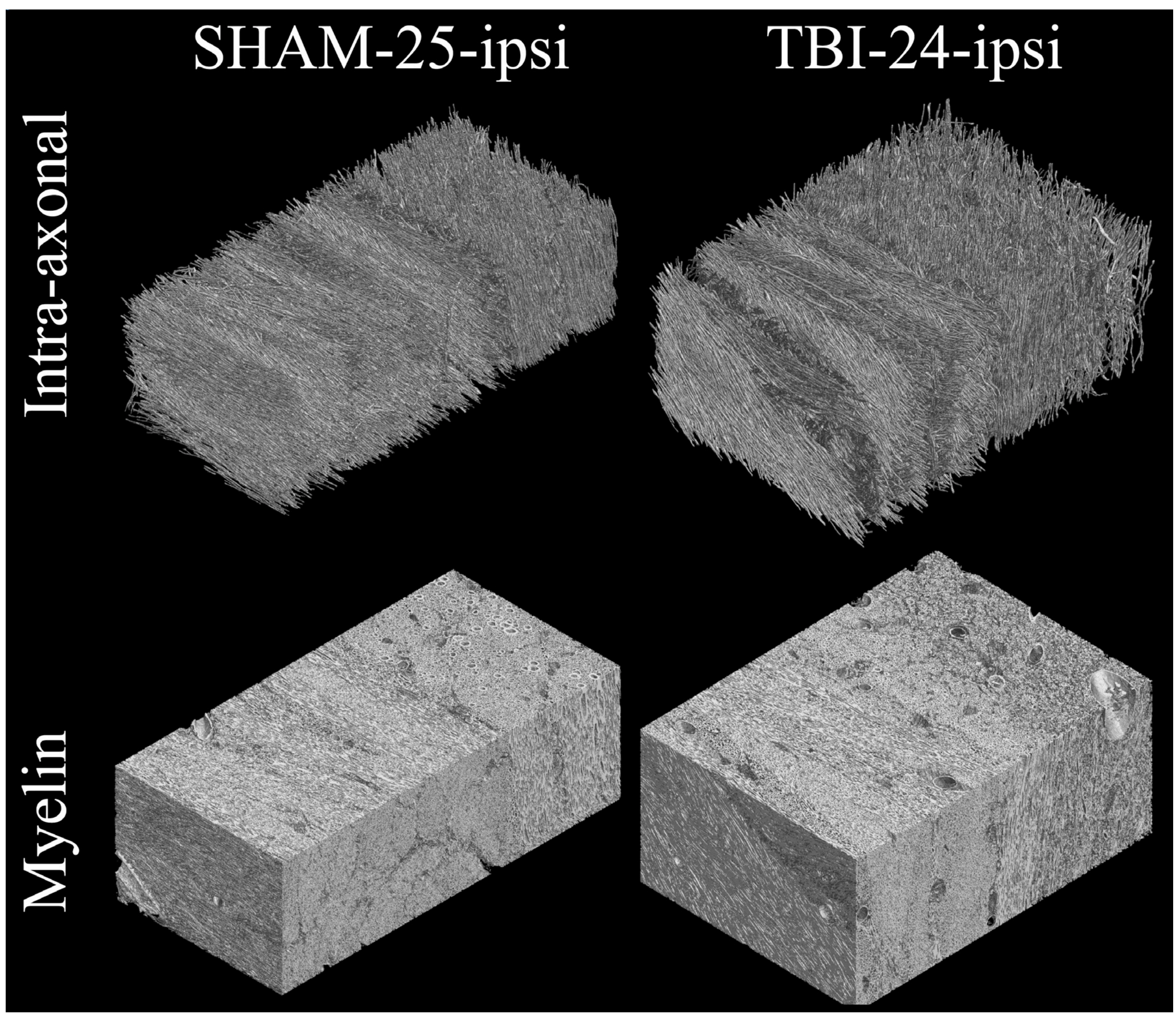

*Figure 3 – Exemplary of in-silico white matter axon phantoms used for Monte-Carlo simulations. Sham rat labelled 25, and TBI rat labelled 24, were used for Monte-Carlo simulations. Labels correspond to the original data*[71,72]*. For each brain sample, the ipsilateral (ipsi) tissue samples are considered. The tissue is extracted from the corpus callosum and cingulum bundles. The intra-axonal spaces are used for the Monte-Carlo simulation of diffusing spins, while the myelin sheaths constitute the magnetizable tissue, perturbing the Larmor frequency of the diffusing spins.*

*Monte-Carlo simulations*

We performed MC simulations in the major fiber bundle of each substrate to simulate an intra-axonal asymmetric spin-echo MRI signal (ASE) $S_{\mathrm{ASE}}(T_E, \Delta T_E)$ with echo time $T_E$ and delayed readout $\Delta T_E$, and an FID signal as a proxy for the gradient echo signal $S_{\mathrm{MGE}}(t)$ with gradient echo time $t$ (here normalized). While details of the simulation framework are described in our previous publication[21], we give here the main details of the simulator: The MC simulations were restricted to intra-axonal spins. Each particle performed random steps drawn from a uniform distribution with a step length of $\delta_l = 0.1\ \mu m$, matching the voxel resolution. The intrinsic diffusivity was set to 2 μm²/ms, corresponding to a time step of $\delta_t = 0.83\ \mu s$. Steps crossing the intra-axonal boundary were rejected. The axon ends were prepared to allow fully periodic boundary conditions within the intra-axonal space, thereby avoiding artificial orientation dispersion. The MC simulations were used to simulate a multi gradient echo signal $S_{\mathrm{MGE}}$ (MGE), and an asymmetric spin echo (ASE) signal $S_{\mathrm{ASE}}$. Signal dephasing caused by the varying Larmor frequency $f(\boldsymbol{r})$, was described by a normalized signal phase tensor $\boldsymbol{\varphi}(T_E; p, t)$ for each random walker $p$, corresponding to each of the six independent tensor elements of the magnetic field tensor $\mathbf{A}(\boldsymbol{r}) \equiv [v \otimes \mathbf{Y}](\boldsymbol{r})$, such that $f(\boldsymbol{r}) = f_0 \widehat{\mathbf{B}}^{\mathrm{T}} \mathbf{A}(\boldsymbol{r}) \widehat{\mathbf{B}}$. Here $v(\boldsymbol{r}) = v_s(\boldsymbol{r})$ or $v_c(\boldsymbol{r})$ and $f_0 = f_s$ or $f_c$, respectively. For each encoded echo time $T_E$: $\boldsymbol{\varphi}(T_E; p, t) = \xi_{T_E}(t) \boldsymbol{\varphi}(p, t-1) + \delta_t \mathbf{A}\left(\boldsymbol{r}_p(t)\right)$. Here the function $\xi_{T_E}(t)$ specifies when a 180-degree RF pulse (here an ideal phase flip) occurs and $\boldsymbol{r}_p(t) = \boldsymbol{r}_{\boldsymbol{p}\mathbf{0}} + \boldsymbol{\Delta}\mathbf{r}_{\mathbf{p}}(t)$ denotes the position of the particle, where $\boldsymbol{r}_{\boldsymbol{p}\mathbf{0}}$ is the initial position and $\boldsymbol{\Delta r}_{\boldsymbol{p}}(t)$ the displacement. Numerically, $\boldsymbol{\varphi}$ was stored as an array with dimensions given by the total number of particles, number of echo times $T_E$ and the 6 tensor components of $\mathbf{A}$. For the MGE signal, $\xi_{T_E}(t) = 1$ for all $t$ and here we write the normalized signal tensor as $\boldsymbol{\varphi}(p; t)$ for simplicity. The ASE signal $S_{\mathrm{ASE}}(T_E + \Delta T_E)$ at a time $T_E + \Delta T_E$, where $\Delta T_E$ denotes the asymmetric delay after $T_E$, were computed for all combinations $2\pi f_0 \widehat{\mathbf{B}}^{\mathrm{T}} \boldsymbol{\varphi}(T_E + \Delta T_E; p) \widehat{\mathbf{B}}$ of $\widehat{\mathbf{B}}$ and $B_0$

$$S_{\mathrm{ASE}}(T_E, \Delta T_E) = \frac{1}{\mathrm{N}} \sum_{\boldsymbol{p}} \mathrm{e}^{-2\pi f_0 \widehat{\mathbf{B}}^{\mathrm{T}} \boldsymbol{\varphi}(T_E + \Delta T_E; p) \widehat{\mathbf{B}}} \quad \text{(ASE)},$$

$$S_{\mathrm{MGE}}(t) = \frac{1}{\mathrm{N}} \sum_{\boldsymbol{p}} \mathrm{e}^{-2\pi f_0 \widehat{\mathbf{B}}^{\mathrm{T}} \boldsymbol{\varphi}(p; t) \widehat{\mathbf{B}}} \quad \text{(MGE)}.$$

(MC simulated MRI signals)

Each simulation used $N = 4 \times 10^6$ random walkers distributed uniformly across the axon bundle, with a computation time of approximately one day per substrate. The simulated signals for each protocol can also be written as

$$S_{ASE}(T_E, \Delta T_E) = \exp\left(-\eta^{SE}(T_E) - \eta^{ASE}(T_E, \Delta T_E) - i\varphi^{ASE}(\Delta T_E)\right)$$

$$S_{MGE}(t) = \exp\left(-\eta^{MGE}(t) - i\varphi^{MGE}(t)\right).$$

Here $\eta^{SE}$, $\eta^{ASE}$, $\eta^{MGE}$ are the sequence-dependent, dimensionless and time-dependent net signal decay functions caused by the heterogenous Larmor frequency shifts. Notice $\eta^{ASE}$ depends also on the echo time $T_E$ [15,37,38]. We calculated the signal within each individual axon in the axonal substrate, where the induced field was generated by *i*) only the axonal sheath microstructure or *ii*) only intra-axonal spherical inclusions packed as in *b*).

The MGE signal was calculated at times $t = 0, 2, 4, \ldots, 18$ ms, while the ASE signal was calculated at $T_E = 20, 22, 24, \ldots, 40$ ms with asymmetric readout $\Delta T_E = 0, 2, 3, \ldots, 20$ ms. Notice that the ASE signal is an SE signal when $\Delta T_E = 0$ ms ($\eta^{ASE}(T_E, 0) = 0$), which meant that we could also extract the SE transverse relaxation from the same simulation.

For the total intra-axonal signal within the major fiber bundle, we estimated the transverse signal relaxation using $\eta^{MGE}(t) = -\ln(|S_{\mathrm{MGE}}(t)|)$, $\eta^{ASE}(\Delta T_E) = -\ln(|S_{\mathrm{ASE}}(\Delta T_E, T_E = 20\ \mathrm{ms})|)$ and $\eta^{SE}(T_E) = -\ln(|S_{\mathrm{SE}}(0, T_E)|)$ for all $\mathbf{B}_0$. Similarly, in order to compare the net signal relaxation to the compartmental signal decay contributions in Eq. (4), we also estimated each compartmental signal decays $\eta_a^{MGE}$, $\eta_a^{ASE}$ and $\eta_a^{SE}$ and phases $\varphi_a^{ASE}$ and $\varphi_a^{MGE}$, where $a$ here denotes different intra-axonal compartments. We chose the lowest simulated echo time $T_E = 20$ ms to maximize estimation accuracy of $\eta^{ASE}$.

First, we investigated if the orientation dependence of $\eta^{SE}, \eta^{ASE}, \eta^{MGE}$ could be described by Eq. (8) for all echo times. Second, we tested if $\eta^{SE}, \eta^{ASE}, \eta^{MGE}$ in a realistic bundle of coherently aligned axons is in fact described by the compartmental signal decays in Eq. (4), and if the scaling on time and $B_0$ is in agreement with short-range structural disorder[47] when $t \gg \tau$. This was done by comparing $\eta^{SE}, \eta^{ASE}, \eta^{MGE}$, at every time and $B_0$ strength, to the computed intra-axonal contributions $\overline{\eta}$ and $\overline{\varphi^2} - \overline{\varphi}^2$ c.f. Eq. (4) across all orientations $\hat{\mathbf{B}}$. Hence, no fitting was involved. Besides confirming if the net signal decay could be captured by Eq. (4), it also enabled us to examine if the time and $B_0$ dependence follows our theoretical prediction: $\overline{\eta} \propto t^{-2v+2}(\gamma B_0 \chi)^2$ for $t \gg \tau$ and $\overline{\varphi^2} - \overline{\varphi}^2 \propto t^2(\gamma B_0 \chi)^2$.

**MRI Imaging**

We also analyzed the orientation dependence of the apparent gradient-echo transverse relaxation rates $R_2^{GE}$ of available datasets. $R_2^{GE}$ relates to $\eta$ when $\eta = tR_2^{GE}$. Although this relation may not hold in general (as explained in the Introduction and Theory sections), it can still be used to assess orientation dependence, provided that the functional form of $\eta$ with respect to time does not vary across orientations. Equation (8) is therefore applicable to describe orientation dependence of $R_2^{GE}$, and the amplitudes $a, b$ and $c$ represent weighted averages of each parameter over all echo times used for estimating $\mathrm{R}_2^{GE}$. As mentioned in the theory section, the offset $c$ includes molecular and macroscopic relaxation.

One dataset was acquired by Sandgaard et al.[20] and contains multi-gradient echo data and dMRI data at 9.4T from rat brain imaged at multiple sample orientations (details can be found in the paper). The second and third datasets examined were acquired by Aggarwal et al.[46] for studying transverse relaxation in fixed human brain stem at 11.7T, and by Denk et al.[16] studying in vivo human brain WM. Transverse relaxation data vs. fiber direction was extracted from Aggarwal et al.[46] and Denk et al.[16] using the online graph reader automeris.io. For the dataset by Sandgaard et al.[20] we fitted the apparent relaxation rate for each sample orientation for every individual voxel in a manually segmented region of Corpus Callosum (CC). The voxel-wise relaxation rates were then fitted to Eq. (8) using the main WM axon orientation determined from the main eigenvector of the fODF scatter matrix (see previous study[20] for details). Using Eq. (8) directly on all data means that we assume an axially symmetric fODF across all voxels, and that we neglect the effect of biological variation across the different voxels (see discussion on limitations and future studies).

# 5| Results

For clarity of presentation, we show here results for two substrates, sham-ipsi-25 and TBI-24-ipsi (cf. Figure 3), as results were comparable across all substrates, including differences between sham and TBI rats.

In supplementary material S1, we show that the Larmor frequency variance from individual axons, both realistic and synthetically grown, respectively. Our simulation shows that none of the simplistic axons are close to the orientation dependence observed for the realistic axons, which means that all structural features are important when modeling relaxation in axons. Second, in supplementary material S2, we show that the Larmor frequency variance acquires a non-negligible contribution from neighboring axons. This means that not only must axons be modeled realistically, but the magnetic field induced by the whole microstructure must be considered.

*Transverse relaxation inside realistic axonal microstructure*

We start by considering the signal decay inside each individual axon, as they describe the three terms in Eq. (4) for ASE and MGE, and the two terms on Eq. (5) for SE. Figure 4-Figure 6 show the “apparent” relaxation rates from intra-compartmental average $\overline{\eta}/t$, and inter-compartmental variance $\left(\overline{\varphi^2} - \overline{\varphi}^2\right)/2t$. We divide by time to make the units 1/s, to make it more intuitive. To keep the figures simple, plots are shown for $f_m = -65$ Hz, $f_s = 328$ Hz and $\theta = 65$ degrees plot across different echo times. Figure S9-Figure S11 in supplementary material show the decay $\overline{\eta}/t$ the other characteristic frequencies. This was chosen as it resembles a mean effect across all simulations. For axons ($i$), we find that $\overline{\eta}/t$ scales approximately as $t^{1/2}$ in agreement with 1D short range structural disorder for all three characteristic frequencies $f_m$. For the spheres (*ii*), the time dependence is also in agreement with 1D short range structural disorder at $f_s = 145$ Hz and $f_s = 328$ Hz as shown in Figure SX in supplementary. At $f_s = 768$ Hz it decreases towards a more linear time dependence, which may be indicative of approaching the static dephasing regime (which was discussed previously on the theory section on time dependence). The phase variance $\left(\overline{\varphi^2} - \overline{\varphi}^2\right)/2t$ scales as $t$ for axons (*i*) and agrees with the spatially averaged Larmor frequency shifts $2\pi^2 t\left(\overline{f^2} - \overline{f}^2\right)$ (green line) estimated directly from the induced magnetic field variance. For the sphere-filled axons (*ii*), the phase variance $\left(\overline{\varphi^2} - \overline{\varphi}^2\right)/2$ agres with $2\pi^2 t\left(\overline{f^2} - \overline{f}^2\right)$ for $f_s = 145$ Hz and $f_s = 328$ Hz, while at $f_s = 768$ Hz, the first order cumulant fails to explain each compartments' signal phase across all echo times.

Our simulations thus confirm the effect of compartmentalization. Namely, these results show that the intercompartmental variance $\overline{\varphi^2} - \overline{\varphi}^2$ in phase shifts can be important when characterizing the net transverse relaxation across multiple compartments as it contributed up to 50% to the total signal decay for axons (cf. Figure 4 and Figure 6).

*Total mesoscopic signal decay*

Having gone through the signal decay from individual axons, we now consider the results for the net signal decay $\eta^{SE}, \eta^{ASE}, \eta^{MGE}$ from all axons.

Figure 7 shows $\eta^{SE}/T_E, \eta^{ASE}/\Delta T_E, \eta^{MGE}/t$ for $f_m = -65$ Hz and $f_s = 328$ Hz at $\theta = 65$ degrees and at different echo times, while Figure S12 in supplementary material show the decay for the remaining characteristic frequencies. We find that the net relaxation decays exhibit a clear time dependence that cannot not be captured by either a purely linear or a purely quadratic function of time, but as a mixture of the terms shown in Figure 4-Figure 6.

Figure S8 plots $\eta^{SE}/T_E, \eta^{ASE}/\Delta T_E, \eta^{MGE}/t$ versus characteristic frequency $f_m$ and $f_s$, respecitvely. The axonal transverse relaxation (*i*) scales as $f_m^2$. This was true across all the simulated echo times. A slower dependence close to 1 was found in the presence of spheres (*ii*) at $f_s = 768$ Hz. This indicates that the net signal decay of axons and intra-axonal mitochondria are well described by the signal's second order cumulant, and that all three relaxation functions scale as $(\gamma B_0 \chi)^2$ - except for the sphere-filled axons with a high characteristic frequency. This observation agrees with previous theoretical work[12] and experimental studies[6,7,59,73] demonstrating that iron-rich tissue should scale linear with $f_s$, when the dephasing gets

Figure 8 shows the net decay of the "apparent" relaxation rates $\eta^{SE}/T_E$, $\eta^{ASE}/\Delta T_E$, $\eta^{MGE}/t$, for a fixed echo time, $f_c = -65$ Hz, $f_s = 328$ Hz and different $\hat{\mathbf{B}}$ orientations, while Figure S13 in supplementary material show the decay for the remaining characteristic frequencies. We fit all the net decays to Eq. (8) (black line) and found that the orientation dependence is well captured by this expression for all characteristic frequencies. The relaxation rate is lowest for SE, and the absolute difference across orientations decreased, as expected, which stems from a slightly reduced intra-compartmental relaxation (cf. Figure 4-Figure 6) and because $\left(\overline{\varphi^2} - \overline{\varphi}^2\right)$ are refocused by the SE pulse sequence.

Using the individual axonal decays, shown in Figure 4-Figure 6, we also compute the expected net decay from Eqs. (4) and (5) (green lines in Figure 8). Here we find that the net signal decays $\eta^{SE}$, $\eta^{ASE}$, $\eta^{MGE}$ can be described by the compartmental cumulant expansion up to second order (i.e. summing over compartments as described in the theory section on compartmentalization) even though the intra-compartmental signal decay $\eta_a$ and phase $\varphi_a$ from within each sphere-filled axons may be influenced by higher order cumulant (i.e. when summing over spins inside each compartment, as described in the theory section on time dependence). This is true for all echo times. A clear difference in orientation dependence can be seen between sham and TBI for the axons in Figure 8. This is indicative by the decay curve across orientation having a minimum at a higher angle for TBI compared to sham. Looking at our simulations in the synthetic axons in supplementary material S1, this difference likely comes from increased beading in the diseased tissue, as demonstrated in a previous study[63].

In general, the rather sparse distribution of strongly magnetized spheres generates a substantially higher signal decay rate (around 10 times higher) than the myelinated axons, even though their volume fraction is 10 times lower than the axons. This shows that intrinsic characteristic frequency (roughly five times higher for the spheres in our simulations) can be more important than volume fraction for transverse relaxation.

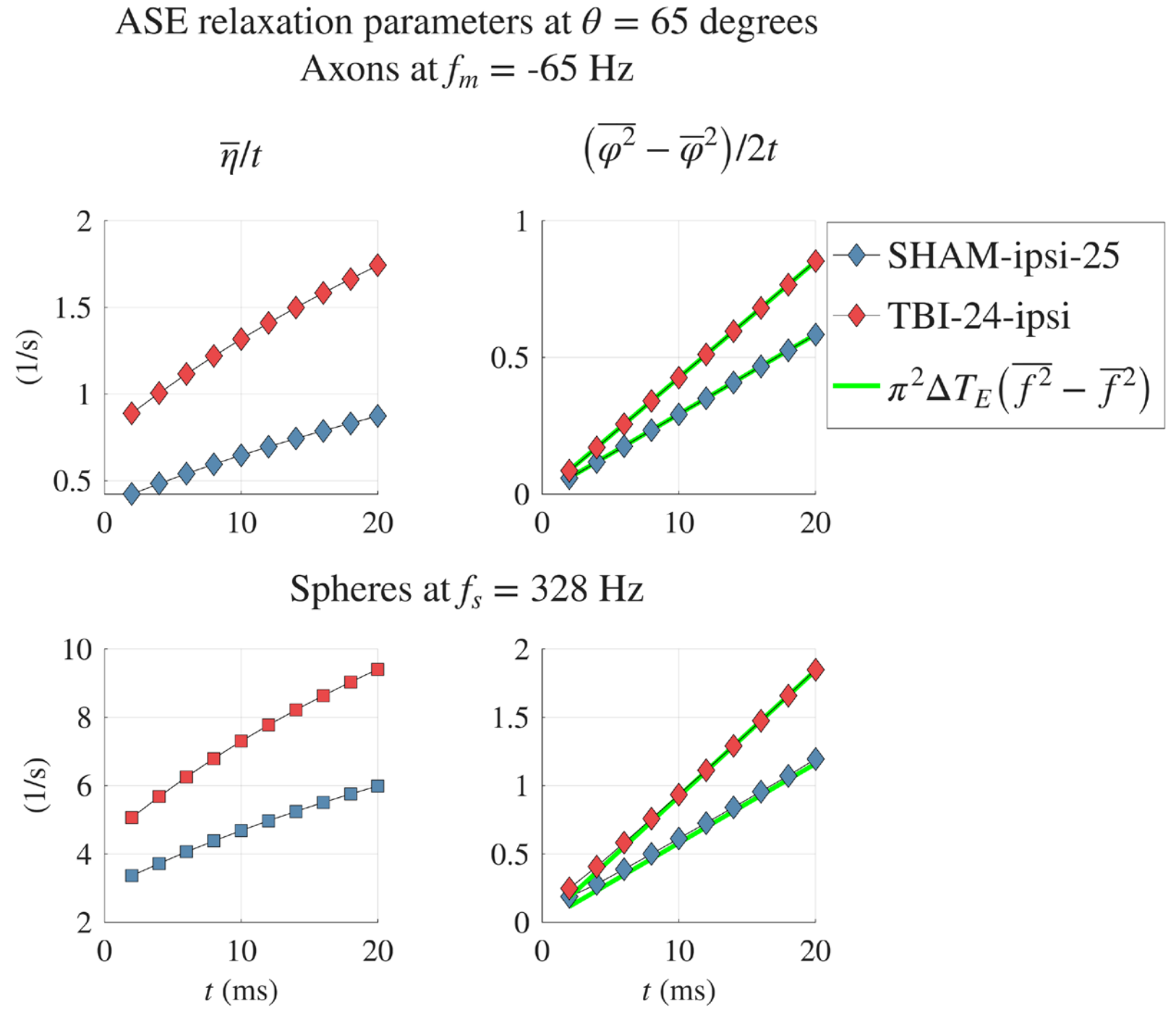

*Figure 4 - Fitting parameters from fitting the ASE signal decay to the magnetic field variances for different echo times $\Delta T_E$ (Eq. (4)). Colors correspond to 2 different axonal substrates. The first row shows the contribution induced by the full EM axonal microstructure, while the bottom for spheres. Relaxation is shown for one characteristic frequency. The green lines in the second column show the estimated phase variance from the induced magnetic field variance $\Delta\boldsymbol{B}$.*

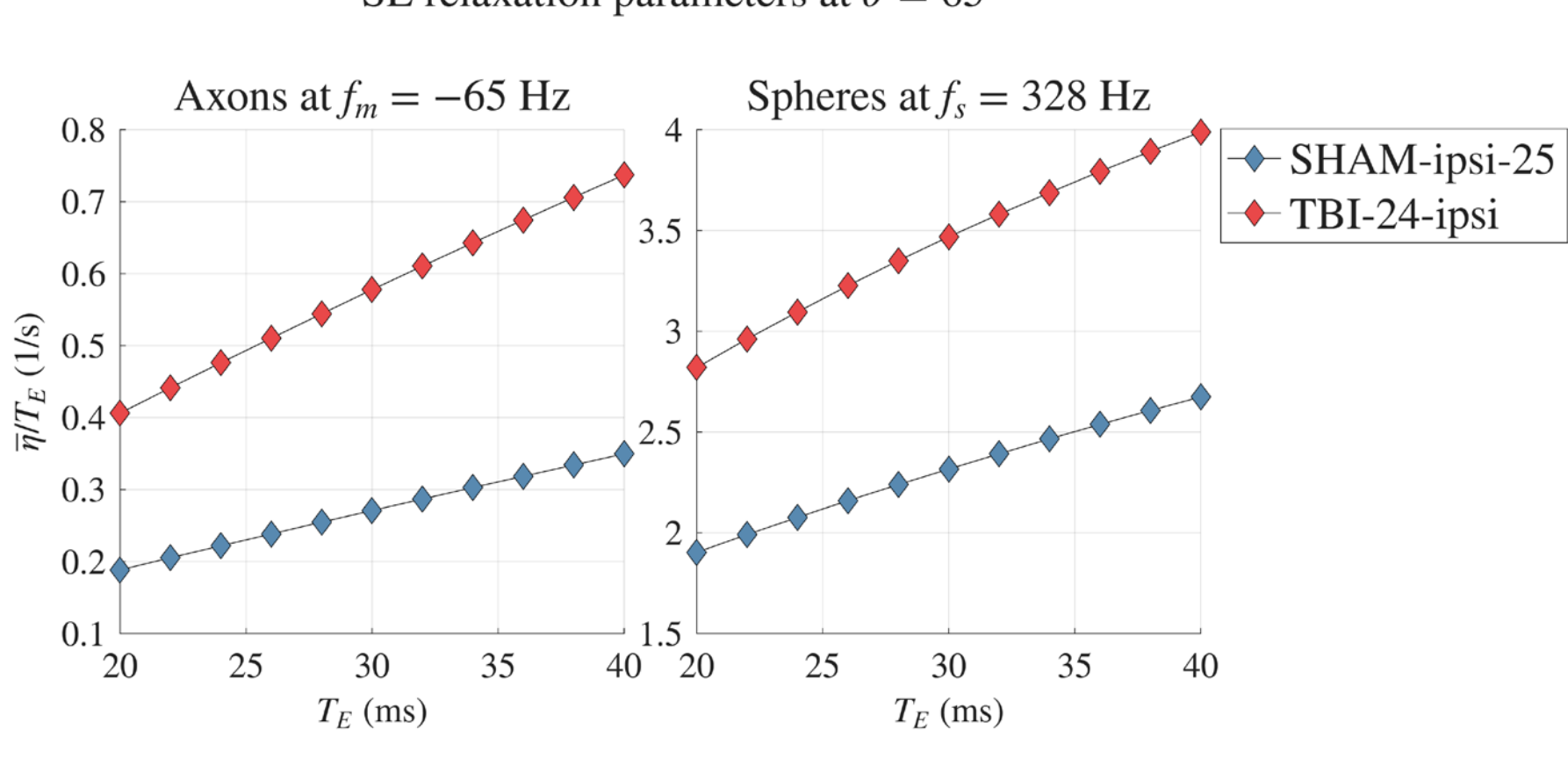

*Figure 5 - Fitting parameters from fitting the SE signal decay to the magnetic field variances for different echo times $T_E$ and $\Delta T_E = 0$, (Eq. (5)). Colors correspond to 2 different axonal substrates. The first image shows the contribution induced by the axonal microstructure, while the second is for sphere-filled axons. Relaxation is shown for one characteristic frequency.*

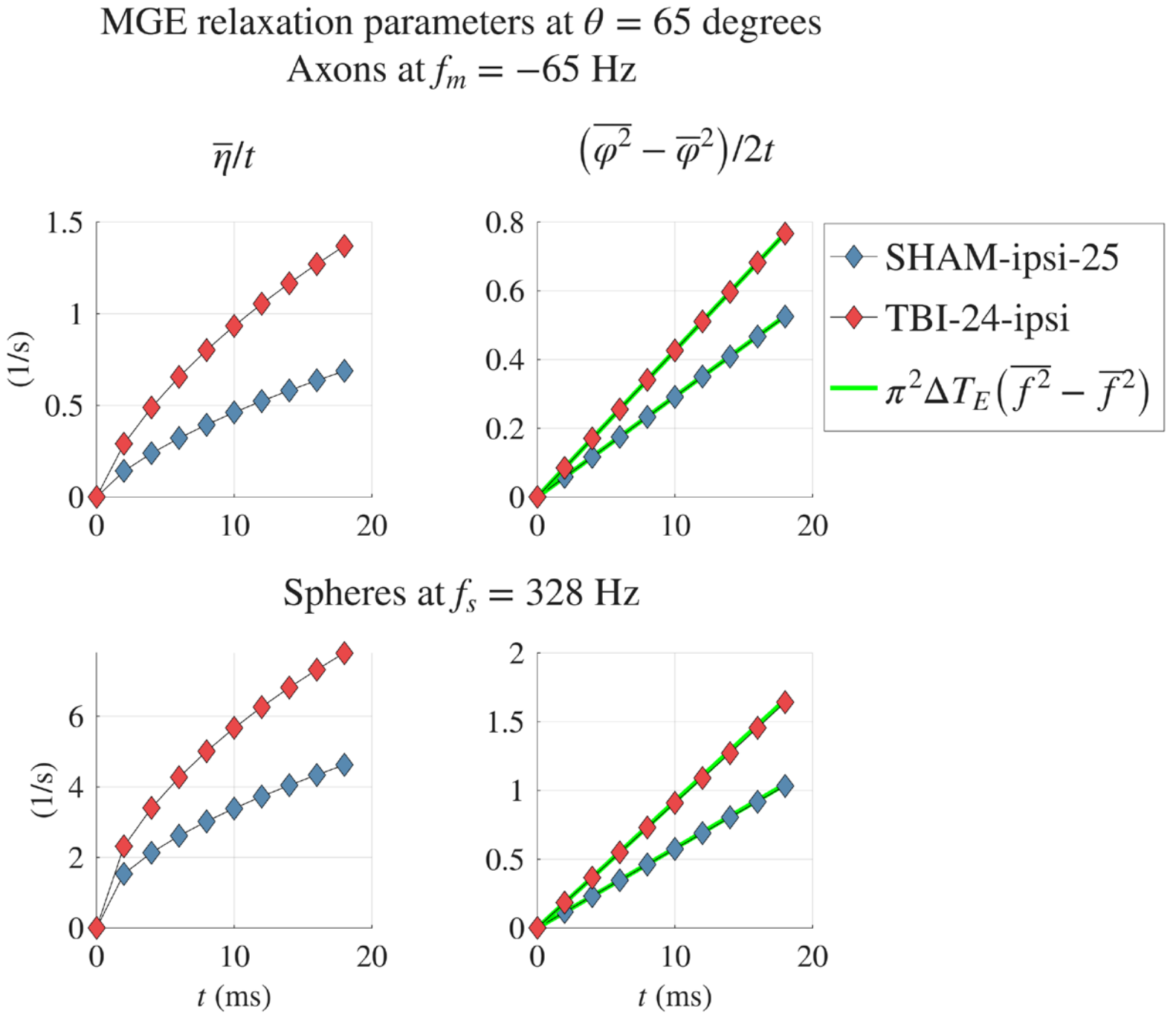

*Figure 6 - Fitting parameters from fitting the MGE signal decay to the magnetic field variances for different echo times* t*, (Eq. (4)). Colors correspond to 2 different axonal substrates. The first row shows the contribution induced by the axonal microstructure, while the bottom is for sphere-filled axons. Relaxation is shown for one characteristic frequency. The green lines in the second column show the estimated phase variance from the induced magnetic field variance* $\Delta \boldsymbol{B}$.

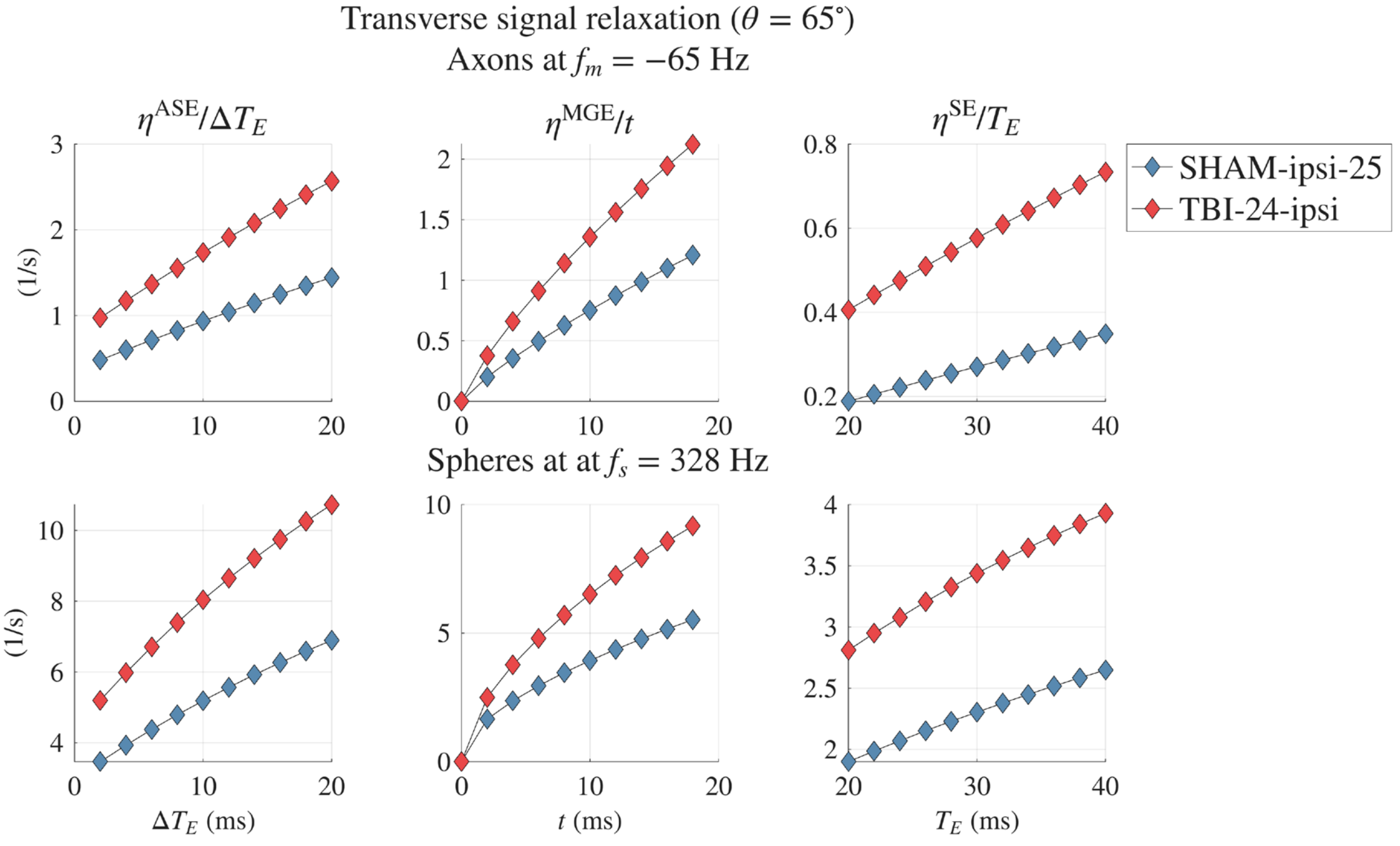

*Figure 7 – Transverse relaxation for an asymmetric spin-echo (ASE) and multi-gradient echo (MGE) signal plotted against echo time. Colors correspond to 2 different axonal substrates. The external field is oriented 65 degrees to the main fiber direction. The first row shows the signal relaxation induced by the unmodified EM axonal microstructure. Relaxation is shown for one characteristic frequency. Bottom row shows the relaxation induced by spheres.*

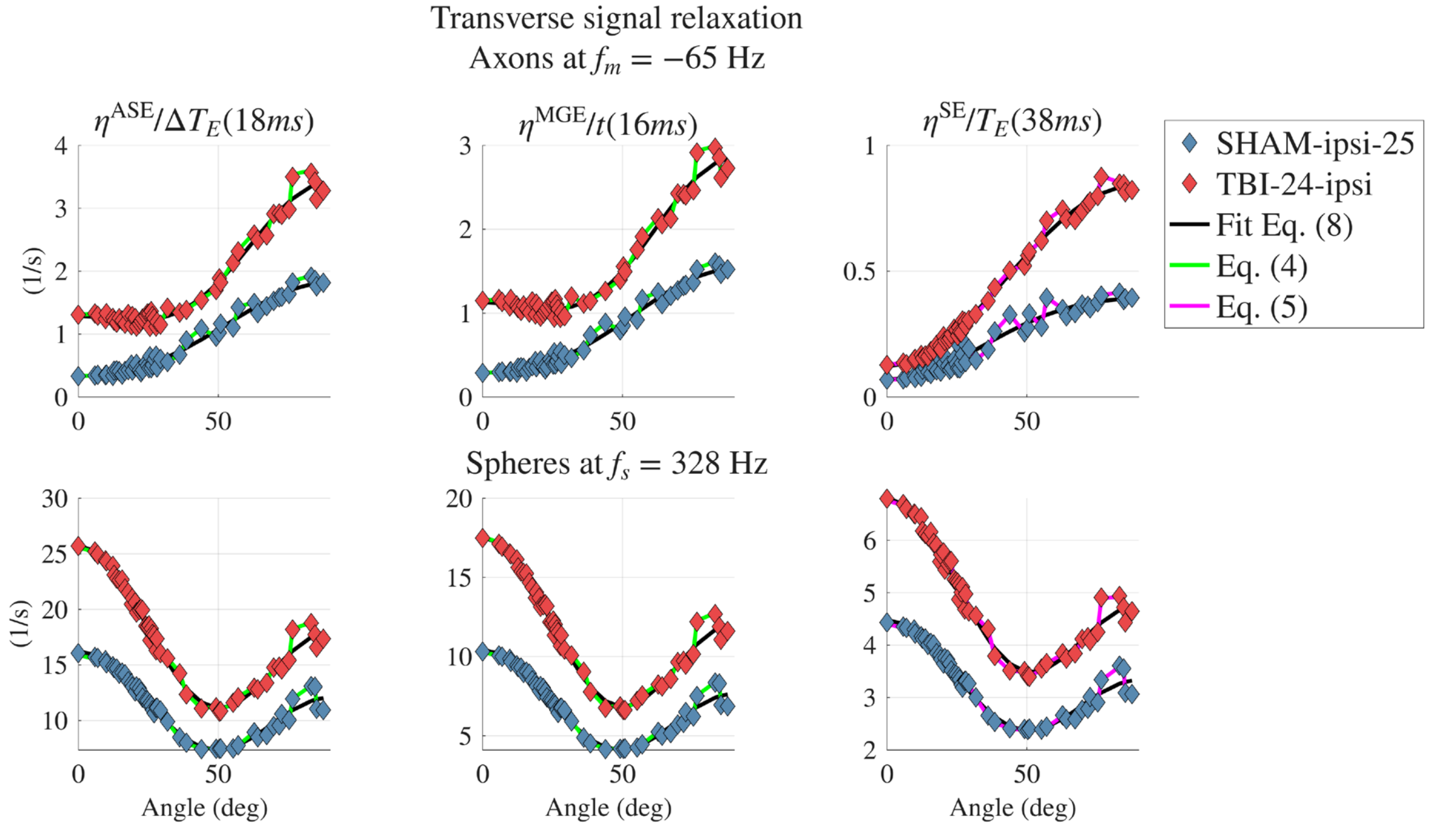

*Figure 8 - Transverse relaxation for spin-echo (SE), asymmetric spin-echo (ASE) and multi-gradient echo (MGE) signals plotted against the angle between the external field and the main fiber bundle at*

*a fixed echo time. Relaxation is shown for one characteristic frequency. Colored points correspond to 2 different axonal substrates. The first row shows the relaxation induced by the realistic axonal microstructure. Bottom row shows the relaxation induced by spheres. Black line shows fitting to Eq. (8) and the green line estimated from Eq. (4) for MGE and ASE and Eq. (5) for SE in magenta.*

**MRI Imaging**

Figure 9 shows the transverse relaxation rate $R_2^{\mathrm{GE}}$ in WM tissue voxels from three different imaging studies. Overall, Eq. (8) could explain the main orientation dependence of all three studies. Table 1 shows the fitting amplitudes. While a true comparison across field strength is hard given the substantial variability arising from biological differences across WM tracts, animal type, and echo times, a clear scaling relation cannot be established at this stage and warrants further investigation. Nevertheless, we still observe that the relaxation rate indeed scales with $B_0$ and the anisotropic amplitudes $a, b$ scales closer to $B_0^2$ than $B_0$.[59] Hence, relaxation anisotropy must be driven by a mechanism that depends on the field strength.

*Table 1 – Amplitudes from fitting MRI data to Eq. (8)*

| **Fitting amplitudes (1/s)** | $B_0$ | Type | Fixation | Tissue | $a$ | $b$ | $c$ |
|---|---|---|---|---|---|---|---|
| Denk et al. | 3T | Human | In vivo | Whole brain | 4.4 | 6.5 | 22 |
| Sandgaard et al. | 9.4T | Rat | Ex vivo | Corpus Callosum | 20 | 27 | 51 |
| Aggarwal et al. | 11.7T | Human | Ex vivo | Transverse Pontine fibers | 33 | 47 | 66 |
| Aggarwal et al. | 11.7T | Human | Ex vivo | Corticospinal tract | 39 | 61 | 74 |

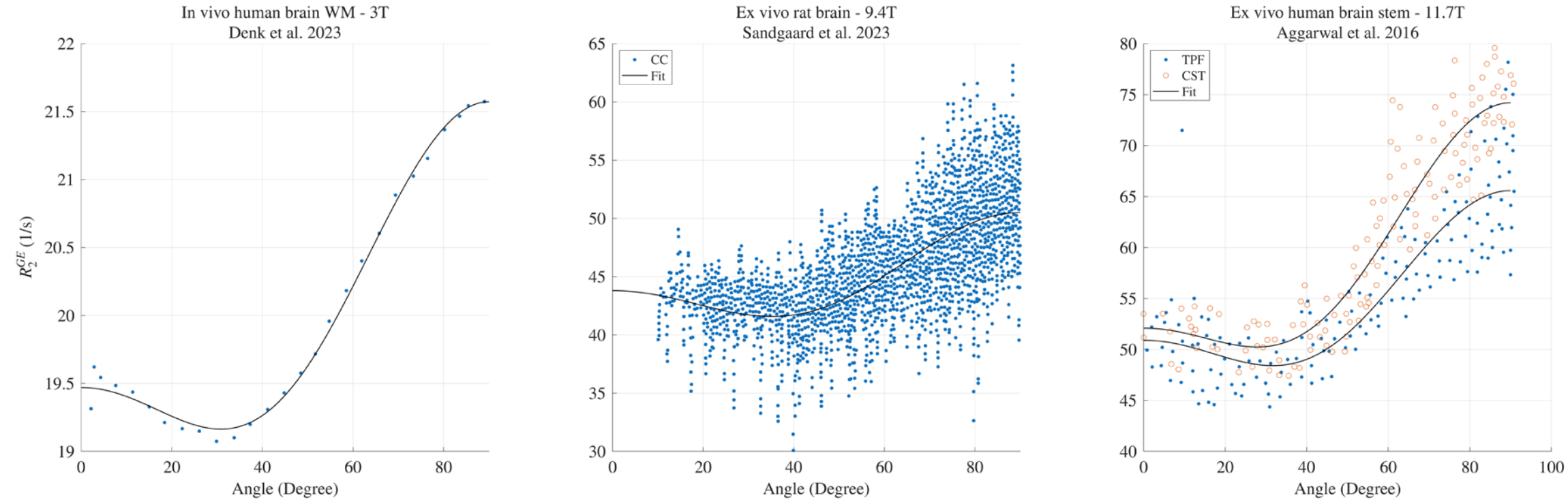

*Figure 9 - Left: Transverse relaxation in in vivo human white matter at 3T. Angles denotes the angle between B0 and the major eigenvector of the diffusion tensor **D** from DTI. Middle: Transverse relaxation in an ROI of corpus callosum from an ex vivo mouse brain imaged at 9.4T. Each point corresponds to a voxel and the angle was found as the angle between B0 and the major eigenvector of the fODF scatter matrix **T**. **T** was estimated using Fiber Ball Imaging[74] (FBI). Right: Transverse relaxation in human brainstem at 11.7T. TPF denotes transverse pontine fibers while CST are corticospinal tracts. Angles was estimated by the angle between B0 and the major eigenvector of the diffusion tensor **D** from DTI..*

# 6| Discussion

## Towards a complete model of WM transverse relaxation rate

*Orientation and time dependence of WM transverse relaxation*

In this work, we studied the orientation dependence of the magnetic field variance and mesoscopic transverse relaxation in realistic white matter magnetic microstructure. We argued theoretically that orientation dependence in an axially symmetric microstructure involves only even order cosines up to 4th order. We also expect our results to remain a good approximation in nearly axially symmetric cases, presumably widespread across the brain. Our Monte-Carlo simulations were also carried out inside axonal bundles, which were part of larger WM substrates (cf. Figure 3) containing multiple bundles oriented with no axial symmetry. This demonstrates that the orientation dependent MR signal relaxation in voxels containing different oriented WM tracts should not be modelled directly using Eq. (8), but instead by using Eqs. (1)-(3) in supplementary material S3, where the fiber orientation distribution (fODF) captures the non-axially symmetric orientation of the WM tracts, and the signal relaxation from a single tract are described by Eq. (8). Hence, our work showed that microstructure alone can account for the orientation dependence previously ascribed to intrinsic susceptibility anisotropy[42]. We validated our results by numerically computing magnetic field variations induced by realistic axonal substrates obtained from electron microscopy, and by additional spherical inclusions added by hand inside the axons to mimic iron-loaded mitochondria[33]. As the characteristic frequency $f_s = \gamma B_0 \chi_s / 2\pi$ of the spheres increased, the weak dephasing regime $\sigma_c \tau \ll 2\pi$ failed to explain the compartmental signal decay (cf. Figure 4-Figure 6). This meant that the second order cumulant expansion *across the intra-compartmental phase distribution* failed in describing the compartmental signal phase $\varphi_c$ and decay $\eta_c$. Nevertheless, our simulations confirmed that the second order cumulant expansion *across compartments* (capturing intercompartmental differences in signal decay and phase) proved valid in describing the net signal, even when the second order cumulant expansion of the intra-axonal signal failed. Properly describing the intra axonal signal phase may therefore require including higher order cumulants, or accounting for strong static dephasing defined as $\sigma_c \tau \gg 2\pi$.[1] This will be investigated in the future.

Besides orientation dependence, we also investigated the time dependence of the total signal decay $\eta$ for different pulse sequences (ASE, SE and MGE) with our Monte-Carlo simulations. Here we found that 1D short-range structural disorder[47] described our findings for axons, while spheres agreed with 1D short-range disorder, when the phase decay could be described in the weak dephasing regime $\sigma_c \tau \ll 2\pi$ . We found that TBI increased the transverse signal relaxation and resulted in a different orientation dependence compared to sham. This difference could be induced by morphological features such as axonal beading (see simulation in Supplementary S1). A recent study[63] using the same axonal substrates found that TBI increases the cross-sectional variance from e.g. enhanced beading, which from the point of view of transverse relaxation would increase the dipolar-like contribution – in agreement with our findings. In Figure S6 we see that TBI-ipsi in C3-C5 (individual axons with different structural features cf. Figure 3) had a higher dipolar modulation, which based on simulations in Supporting material S1 indicates a higher axonal sinuosity, in agreement with the previously mentioned study[63]. Compared to the structurally modified axons, C2 to C6, indicates that beading along the axis is more important than non-circular cross-sections, as the magnetic field variance changed the most when beading was removed.

*Scaling of magnetic susceptibility in WM signal*

The transverse signal relaxation $\eta$ from myelinated axons was in good agreement with the prediction from the second signal cumulant, scaling with field strength to the second power, as seen in Figure 8. This means that transverse relaxation from myelinated axons in our simulations should scale with $\zeta_m \chi_m^2$ for the considered range of $f_m = \gamma \chi_m B_0 / 2\pi$ values, which we believe encompasses a wide span of realistic conditions. For spherical perturbers, neither a linear nor a squared scaling relation of magnetic susceptibility was adequate for explaining all the simulations across all the substrates. In practice we found from the $f_s = [140,\ 328]$ Hz simulations that when $f_s$ was sufficiently low, which is likely comparable to the iron containing cells in WM[69,70], the transverse relaxation scaled approximately as $f_s^2$, and thus as $\zeta_s \chi_s^2$. As the characteristic frequency increases, the exponent of the scaling decreases. We expect that the $f_s^2$ scaling observed at lower characteristic frequencies will eventually transition toward a $f_s$ linear relationship at higher characteristic frequencies. This shift signifies the transition into the strong static dephasing regime[12,38], where the standard cumulant expansion fails and a non-perturbative approach is required to accurately describe the mesoscopic signal decay. Such linear $f_s$ scaling is typically expected in iron-rich GM, where the high local field variance around iron deposits drives the system out of the weak dephasing regime. In both dephasing regimes, the decay mesoscopic $\eta$ is expected to scale linear with the concentration $\zeta_s$. This means that the apparent scaling exponent $f_s$ depends on the experimental setup and iron content. It is therefore important to match the correct model to the experimental design, and the best model may be different in iron-deficient WM such as the Corpus Callosum compared to superficial WM[6,7,69]. In practice, estimating and distinguishing such scaling exponents of magnetic-susceptibility–induced relaxation is challenging, not only because identifying the correct power law in time is inherently difficult, but also due to practical limitations such as the need for several orders of magnitude in dynamic range, and a strong dependence on the earliest time points. Identifying the correct power law may therefore be hard with finite and noisy data sets and further compounded by the presence of biological variability.

## Limitations

### *Contribution from point-like spherical sources*

In our simulations, spherical inclusions had a radius comparable to the axons, as smaller spheres gave numerical errors in the computed magnetic fields. Hence, we could not simulate the effect of strongly magnetized point-like particles. For example, ferritin molecules have a diameter around 4 nm and magnetic susceptibility 520 ppm[70]. In comparison, our resolution was 0.1 μm in the axonal substrates. A voxel containing one ferritin molecule would thus have an effective susceptibility around 140 ppb. This means that our simulation resolution prevented us from probing strong frequency shifts induced by point-like particles. Instead, we focused on larger spheres to mimic iron-containing cells. Another limitation of this study is that we only packed intra-axonal spheres and simulated intra-axonal Monte Carlo signals. We modeled iron sources as spheres within the axons based on evidence from Meguro et al.[33], which demonstrates that mitochondria contain iron and possess an effective radius of approximately 0.5 μm. While Meguro et al. also showed iron deposited outside the axons (mainly contained in neuroglia), this was not directly implemented in our simulations. However, since our simulations focus on intra-axonal water, we believe our approach is sufficient for the current study. From a magnetic field perspective, the field inside a single axon is caused by iron-containing spheres elsewhere in the tissue can barely distinguish whether those spheres are located in neighboring axons or in the extra-axonal space. Hence, we believe it is sufficient to pack spheres only in the intra-axonal space to give insights into the effect of iron on the intra-axonal signal decay. This approach made the packing of the spheres more feasible and avoided the need for speculative segmentation of the extra-axonal space, while still capturing the representative magnetic environment. Simulating the extra-axonal signal will however be considered in future studies. But, based on our investigation of the Larmor frequency variance (cf. Figure S2Figure S6 and Figure S7), which was larger in the extra-axonal space across all substrates, we expect the transverse relaxation rate to be faster outside than inside the axons, and that extra-axonal spheres would give a rise to similar time- and orientation-dependents effects, if packed with similar volume fraction.

*Beyond the modelled orientation dependence*

While our results are especially applicable to ex vivo tissue, the situation in vivo is more complex: Transverse relaxation arises not only from magnetic field variations induced by the magnetized microstructure, but also from additional mechanisms. For instance, molecular dipole-dipole relaxation $\eta^{Mol}$ contributes both ex vivo and in vivo, is most likely orientation independent, and increases linearly with time[1]. Such relaxation will contribute to the orientation-independent amplitude $c$ in Eq. (8). Recent work has nevertheless indicated the presence of orientation-dependent relaxation[34], arising from dipole–dipole interactions between water and the myelin sheath – the so-called THB model. The THB model explains orientation-dependent relaxation as arising when a water proton transiently forms a hydrogen bond with an ordered (anisotropic) macromolecular proton pool. If the interaction strength is sufficiently high, the relaxation becomes largely independent of $B_0$ strength. As an example, provided by the authors: assume that a sufficient fraction of water molecules participates in these events (around 10%) at every moment. If a proton remains in the transiently bound state for an average lifetime of approximately 10 to 60 ns, during which it experiences a strong dipolar interaction around 20 $(\mathrm{s} \cdot \mathrm{ns})^{-1}$, the net relaxation rate ranges between 1-4 $1/s$ to 10-35 $1/s$, respectively, depending on the orientation but independent of field strength commonly used in MRI[34]. The minimum relaxation rate will be around 42 degrees. Disentangling the respective contributions from THB and microstructure in realistic white matter will require carefully designed experiments capable of isolating each effect, which we aim to pursue in future work.

On a microscopic scale, cardiac pulsations may further introduce time- and orientation-dependent transverse relaxation, modulated by blood vessel size. In line with these multiple contributing factors, recent work in gray matter has demonstrated that it may be possible to disentangle heme and non-heme relaxation[75], suggesting that a similar approach could help to separate different sources of transverse relaxation in WM as well.

*Experimental discrepancies*

The proposed orientation dependence of the transverse relaxation rate could describe the experimental findings investigated. However, other in-vivo studies in brain WM, using the principal eigenvector of the diffusion tensor as a proxy for the orientation of the axonal microstructure, found that the relaxation rate had a large dipolar contribution but with a minimum rate around 20 degrees compared to 30 degrees in the studies considered here (cf. Figure 9). This exact combination of angle minima and dipolar-like modulation cannot be explained by Eq. (8). But, Denk et al. [19] found that venous blood vessel do not necessarily follow the direction of the axons, and this can introduce an offset in the relaxation rate angular dependence. Denk et al. also demonstrated that anisotropic voxel shapes can elevate the effect of venous blood on transverse relaxation. This is why we limited our experimental comparison to studies with isotropic voxel resolution. Hence, we do not see this discrepancy as a disagreement with our work. In addition, introducing an angle-offset to Eq. (8), we can describe said studies[19] with such a lower angle minimum. We also emphasize that a voxel's signal includes convolution with a fiber orientation distribution due to orientation dispersion. In human WM, orientation dispersion is ever-present[51,58] and can include crossing fibers[76] etc. and such effects should not be neglected. Eq. (8) should therefore only be used directly to describe the voxel transverse relaxation rate when the fODF is axially symmetric.

A limitation of the present MRI datasets is that multiple voxels were used to examine orientation dependence. Since WM voxels may differ biologically across tracts, this introduces potential bias. A more rigorous assessment would require a carefully designed experiment on the same tissue sample, acquired at identical echo times but under different orientations and field strengths, to fully avoid confounding tissue variations. Such an experiment could provide clearer insights into both orientation and time dependence and will be carried out in the future.

**Propositions for transverse relaxation modelling and susceptibility estimation**

Quantitative Susceptibility Mapping (QSM) has been combined with models of transverse relaxation to disentangle dia- and paramagnetic contributions from e.g. myelin and iron in the brain[39,40,75]. Our results here challenge the interpretability of susceptibility values obtained from susceptibility models combining phase and transverse relaxation, and such models should be used with caution. This is because an assumption in such models is that both sources (myelin and iron) contribute to the MR signal under a static dephasing regime, particularly at long echo times where the signal can no longer be accurately described by a Taylor expansion. Under this assumption, transverse relaxation is expected to scale linearly with the bulk susceptibility $\overline{\chi}$ and thus with myelin and iron concentration - and with the main magnetic field $B_0$. However, our simulations did not support the presence of a strong static dephasing regime for myelinated axons; instead, the transverse relaxation scaled non-linearly with time, corresponding to short-range structural disorder. As the transverse relaxation for myelin scaled as $B_0^2$, it must also scale as $\chi^2$ with the intrinsic magnetic susceptibility, as varying $B_0$ is indistinguisable from varying $\chi$ (cf. Eq. (1)). Moreover, axons induced substantially lower transverse relaxation rates than spherical inclusions in our simulations, despite spheres having a lower bulk susceptibility, which may complicate the estimation of myelin content in the presence of non-negligible relaxation induced by iron. For spheres, the relaxation rate did not conform strictly to either a linear or quadratic scaling, but somewhere in between. Even if an appropriate scaling relationship were identified for each source, the absolute relaxation rates remain highly dependent on microstructural morphology, and orientation dependent higher-order correlations between the two magnetic sources are also present which may lead to additional complexity in describing the transverse relaxation.

Hence, our work highlights a major challenge of using the transverse relaxation rate directly for susceptibility estimation. Nevertheless, it demonstrates that time and orientation independent relaxation parameters can still be identified and that these reflect underlying morphology and magnetic properties, thereby potentially providing biomarkers for neurodegenerative diseases. In this respect, our results agree with Winther et al.[41] findings on the effect of axonal morphology on transverse relaxation. Here we extended their results by showing that the whole microstructure must be considered when describing the signal's transverse relaxation.

Overall, we propose to describe the measured MRI signal similarly to the Standard Model of WM[77], where a mesoscopic signal kernel is convolved with a fiber orientation distribution function (fODF). Our model may be unified with SM such that diffusion weighting can be used to disentangle signals from the intra- and extra-axonal spaces, but also to probe the orientation dependence of both transverse relaxation and Larmor frequency shifts without sampling the MRI at multiple $B_0$ orientations[78]. Merging diffusion and susceptibility modelling, to improve parameter estimating, is an on-going study[78] and will be presented in the future.

# 7| Conclusion

We demonstrated that realistic white matter substrates containing myelinated axons and spherical inclusions induce orientation- and time-dependent transverse relaxation. Signal dephasing caused by myelinated axons follows a quadratic dependence on susceptibility and field strength, whereas spherical inclusions, such as iron-rich neuroglia, can produce strong relaxation effects that deviate from this simple power-law. However, the time-dependent power-law signature arising from structural disorder in white matter is weak and may be difficult to detect at currently achievable noise levels, echo times, and field strengths, because the power-law curvature over typical echo times deviates only slightly from a linear trend. Transverse relaxation varied by several Hz depending on orientation and $B_0$ field, making it more feasible to measure, as demonstrated with actual MRI data.

These findings challenge current models for estimating magnetic susceptibility from multiple sources and highlight the importance of accounting for both microstructural geometry and compartment-specific magnetic properties in modeling transverse relaxation. Incorporating both orientation-dependent effects may provide more robust biomarkers of tissue pathology, less biased by experimental conditions. Future work will focus on extending this framework to include diffusion-weighted imaging, with the goal of enabling more accurate in vivo characterization of white matter microstructure and its alterations in disease.

## 8| Funding

This study is funded by the Independent Research Fund Denmark (Grant Number 10.46540/3103-00144B).

## 9| Competing interests

The author(s) declare no competing interests.

## 10| Author Contributions

**Anders Dyhr Sandgaard:** Conceptualization, Methodology, Software, Formal analysis, Investigation, Writing - Original Draft, Visualization.

**Rafael Neto Henriques**: Methodology, Writing - Review and Editing,

**Noam Shemesh**: Methodology, Resources, Writing - Review and Editing,

**Sune Nørhøj Jespersen**: Conceptualization, Methodology, Writing - Review & Editing, Supervision, Project administration, Funding Acquisition

## 11| Availability of data

The code used to process the data, calculate the magnetic fields, and perform MC simulations is available upon reasonable request to the corresponding author.

## **12|** Symbols

$\hat{\boldsymbol{n}}$: Cylinder direction vector

$\mathbf{T}$: Mean orientation tensor of cylinder directions $\hat{\boldsymbol{n}}$. Relates to second moment of $\mathcal{P}\,(\hat{\boldsymbol{n}})$

$\chi$: Magnetic Susceptibility

$\mathbf{B}_0 = \hat{\mathbf{B}} B_0$: Magnetic field vector with direction $\hat{\mathbf{B}}$ and magnitude $B_0$

$\theta$: Angle between $\hat{\mathbf{B}}$ and $\hat{\boldsymbol{n}}$

$\xi(t)$: Spin-flip function describing the signal encoding protocol

$V$: Macroscopic (voxel) volume

$\mathcal{M}$: Mesoscopic volume

$l, \tau, D$: Correlation length of microstructure, correlation time $\tau = l^2/D$, where $D$ is diffusivity

$\nu = p + d$: Dynamical exponent of microstrucutre, where $p$ denotes the structural disorder class and $d$ the effective dimension of diffusion process.

$\eta(t)$: Transverse relaxation decay function of signal

$\eta^{Mol}(t)$: Molecular transverse relaxation decay function of the signal

$\eta^{Meso}(t)$: Mesoscopic transverse relaxation decay function of signal

$\eta^{Macro}(t)$: Macroscopic transverse relaxation decay function of signal

$\gamma$: Gyromagnetic ratio of water.

$\mathbf{Y}(\boldsymbol{r})$: Dipole field tensor

$\Delta\mathbf{B}(\boldsymbol{r})$**:** Induced magnetic field of tissue

$f(\boldsymbol{r})$: Local Larmor frequency shift induced by $\Delta\mathbf{B}(\boldsymbol{r})$

$\varphi(t)$: Signal phase at time t induced by $\Delta\mathbf{B}(\boldsymbol{r_t})$

$w_c$: Signal fraction of compartment *c* in volume $\mathcal{M}$

$f_c$: Intra-compartmental mean Larmor frequency shift $\Omega(\boldsymbol{r})$ in compartment *c*

$\sigma_c^2$: Intra-compartmental variance of Larmor frequency shift $f(\boldsymbol{r})$ in compartment *c*

$\langle \dots \rangle$*:* Averaging across spins

$\overline{(\dots)} = \sum_c w_c\,(\dots)$: Averaging across compartments *c* with signal fraction $w_c$

$\overline{\Omega} = \sum_c w_c\,\Omega_c$: Inter-compartmental mean of intracompartmental mean Larmor frequency shift $\Omega_c$

$\overline{\sigma^2} = \sum_c w_c\,\sigma_c$: Inter-compartmental mean of intracompartmental variance of Larmor frequency shift

$\overline{\varphi} = \sum_c w_c\,\varphi_c$: Inter-compartmental mean of intracompartmental phase $\varphi_c$

$\overline{f^2} - \overline{f}^2$: Inter-compartmental variance of Mean Larmor frequency shift $f_c$

$\overline{\varphi^2} - \overline{\varphi}^2$*:* Inter-compartmental variance of intra-compartmental phases $\varphi_c$

$a(t), b(t), c(t)$: Fitting amplitudes of $\eta(t)$ versus the angle $\theta$

$v(\boldsymbol{r})$: Spatial indicator function

$\zeta$: Volume fraction of $v$ in volume $\mathcal{M}$

$t$*:* gradient echo time

$T_E$: Spin echo time

$\Delta T_E$: Delayed time after $T_E$ of signal read-out

# 13| Supplementary Materials captions

*Figure S1 - Overview of synthetic axons filled with randomly packed spheres. A tube with randomly varying radius was generated and each cross-section of the tube was shifted randomly in the radial direction by an amount* $L_{shift}$. *First two rows show the tube filtered by a 3D Gaussian filter, where the first two collumns show the tube before smoothing, and the last two rows after smoothing. First row shows without any shift* $L_{shift}$, *while the second shows with shift. The second and third rows show axons with 2D Gaussian smoothing perpendicular to the axon, and the latter two rows with 1D smothing longitudinally.*

*Figure S2 - Larmor frequency variance indcued by synthetic axonal myelin sheath with scalar susceptibility. First six rows show the intra-axonal magnetic field variance for 3D, 2D or 1D Gaussian smoothing and cross-sectional shift* $L_{shift}$. *The last six rows show variance outside the synthetic axons. X-axis denotes the angle between B0 and axon.*

*Figure S3 - Larmor frequency variance induced by randomly packed spheres with scalar susceptibility inside a synthetic axon. First six rows show the intra-axonal magnetic field variance for 3D, 2D or 1D Gaussian smoothing and cross-sectional shift* $L_{shift}$. *The last six rows show variance outside the synthetic axons. X-axis denotes the angle between B0 and axon.*

*Figure S4 - Parameters of Eq. (8) after fitting Larmor frequency variance induced inside synthetic axon generated by either the synthetic axons or by randomly packed sphered in the axons. Here is shown for a radius* $R_S/R_0 = 0.25$ *for the spheres compared to the mean axon radius. X-axis denoted the amount of cross-section shifts induced for each axon slice, while the y-xais the size of the smoothing filter. First two rows show for 3D Gaussian smoothing, the next two 2D Gaussian smoothing and the latter rows 1D Gaussian smoothing. Colors are clippped in order to visualize the whole range of values.*

*Figure S5 - In-silico white matter axon phantoms used for Monte-Carlo simulations. Eight different substrates from two different sham rats labelled 25 and 49 and two different TBI rats labelled 2 and 24 were used for Monte-Carlo simulations. Labels correspond to the original data. For each brain, both ipsilateral (ipsi) and contralateral (contra) tissue samples are considered. The tissue is extracted*

*from the corpus callosum and cingulum bundles. The intra-axonal spaces are used for the Monte-Carlo simulation of diffusing spins, while the myelin sheaths constitute the magnetizable tissue, perturbing the Larmor frequency of the diffusing spins.*
*For each substrate, we considered the magnetic field variance induced by each axon. In the spirit of Winther et al.*[41] *we synthesized 6 different axons with varying microstructural features to investigate the intra-axonal and extra-axonal magnetic field average and variance. Labeling was kept as in Winther et al. for consistency.*

*Figure S6– Larmor frequency variance induced by individual axonal myelin sheaths or intra-axonal spherical sources (C5 spheres) versus the angle between B0 and average direction of the axon. First three rows show the variance inside every axon, while the last row shows extra-axonal variance. Labels (C2-C6) indicate the diffent morphological features considered (see Figure 3). Each line shows results for 2 different WM substrates, including fits and minima.*

*Figure S7 - Larmor frequency variance, induced by axonal myelin sheaths with scalar susceptibility, versus angle between B0 and the average direction of the major fiber bundle. First row shows the intra-axonal variance for 2 different WM substrates, with the first column displays the variance in the extra-axonal space. Second row shows the variance by intra-axonal spherical sources with scalar susceptibility.*

*Figure S8 - Transverse relaxation for spin-echo (SE), asymmetric spin-echo (ASE) and multi-gradient echo (MGE) signals plotted against the* $B_0$ *strength squared, and angled at 65 degrees to the axon bundle. Colored points correspond to 2 different axonal substrates. The first row shows the relaxation induced by the realistic axonal microstructure. Bottom row shows the relaxation induced by spheres.*

*Figure S9 - Fitting parameters from fitting the ASE signal decay to the magnetic field variances for different echo times* $\Delta T_E$ *(Eq. (4)). Colors correspond to 2 different axonal substrates. The first row shows the contribution induced by the full EM axonal microstructure, while the bottom for spheres. Relaxation is shown for one characteristic frequency. The green lines in the third column show the estimated phase variance from the induced magnetic field variance* $\Delta\boldsymbol{B}$.

*Figure S10 - Fitting parameters from fitting the MGE signal decay to the magnetic field variances for different echo times* $t$*, (Eq. (4)). Colors correspond to 2 different axonal substrates. The first row shows the contribution induced by the axonal microstructure, while the bottom is for sphere-filled axons. Relaxation is shown for one characteristic frequency. The green lines in the third column show the estimated phase variance from the induced magnetic field variance* $\Delta\boldsymbol{B}$.

*Figure S11 - Fitting parameters from fitting the SE signal decay to the magnetic field variances for different echo times $T_E$ and $\Delta T_E = 0$, (Eq. (5)). Colors correspond to 2 different axonal substrates. The first row shows the contribution induced by the axonal microstructure, while the bottom is for sphere-filled axons. Relaxation is shown for one characteristic frequency.*

*Figure S12 – Transverse relaxation for an asymmetric spin-echo (ASE) and multi-gradient echo (MGE) signal plotted against echo time. Colors correspond to 2 different axonal substrates. The external field is oriented 65 degrees to the main fiber direction. The first row shows the signal relaxation induced by the unmodified EM axonal microstructure. Relaxation is shown for one characteristic frequency. Bottom row shows the relaxation induced by spheres.*

Figure S13 - Transverse relaxation for spin-echo (SE), asymmetric spin-echo (ASE) and multi-gradient echo (MGE) signals plotted against the angle between the external field and the main fiber bundle at a fixed echo time. Relaxation is shown for one characteristic frequency. Colored points correspond to 2 different axonal substrates. The first row shows the relaxation induced by the realistic axonal microstructure. Bottom row shows the relaxation induced by spheres. Black line shows fitting to Eq. (8) and the green line estimated from Eq. (4) for MGE and ASE and Eq. (5) for SE in magenta.

# 14| References

# Supplementary Materials

## Axonal microstructure and compartmentalization impact the orientation and time dependence of mesoscopic transverse relaxation

**Anders Dyhr Sandgaard[1], Rafael Neto Henriques[2], Noam Shemesh[2], Sune Nørhøj Jespersen[1,3]**

[1]Center of Functionally Integrative Neuroscience, Department of Clinical Medicine, Aarhus University, Denmark

[2]Champalimaud Research, Champalimaud Centre for the Unknown, Lisbon, Portugal

[3]Department of Physics and Astronomy, Aarhus University, Denmark

# S1 - Larmor frequency variance inside hollow synthetic axons packed with spheres

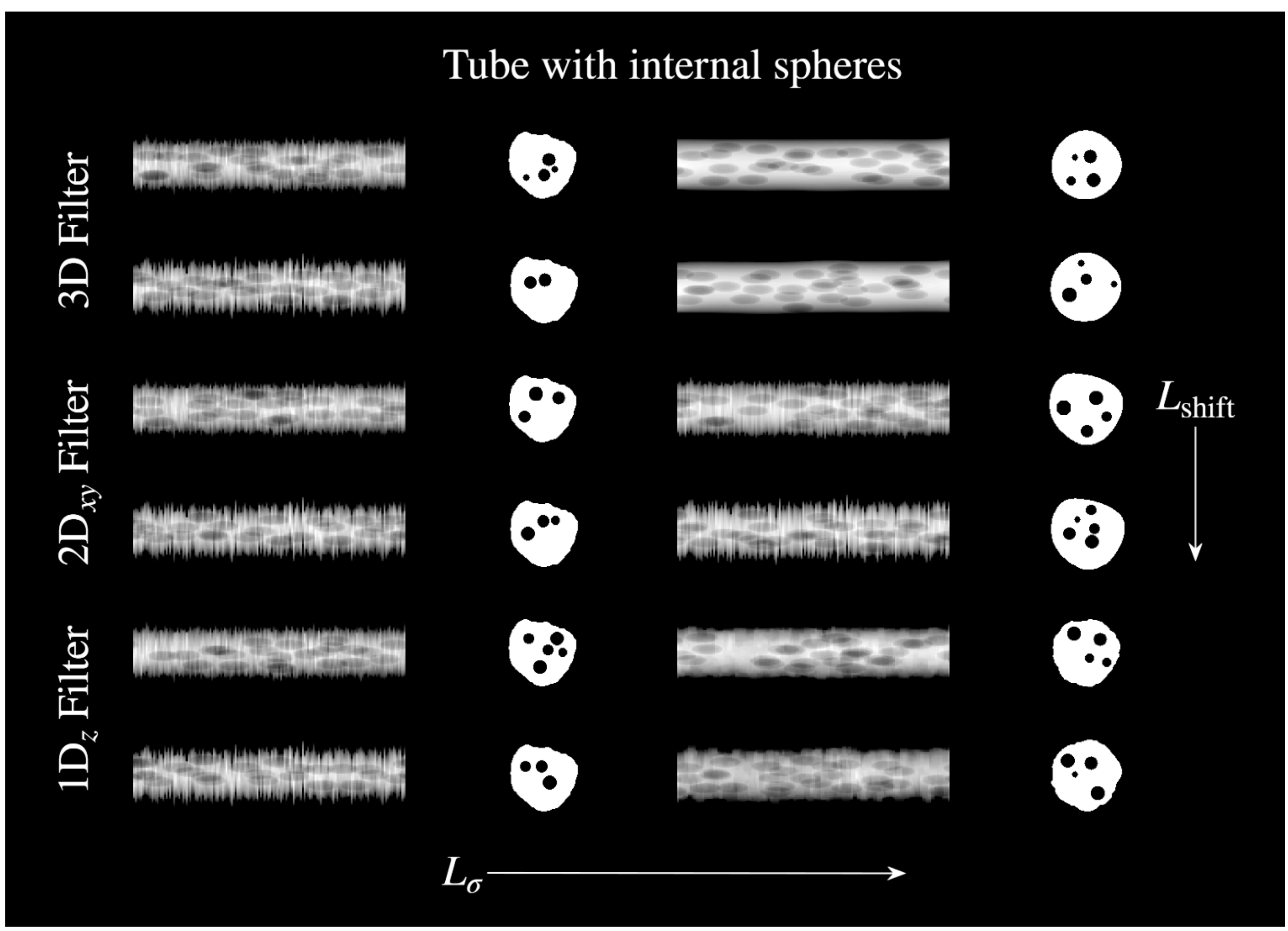

Figure S1 - Overview of synthetic axons filled with randomly packed spheres. A tube with randomly varying radius was generated and each cross-section of the tube was shifted randomly in the radial direction by an amount $L_{shift}$. First two rows show the tube filtered by a 3D Gaussian filter, where the first two collumns show the tube before smoothing, and the last two rows after smoothing. First row shows without any shift $L_{shift}$, while the second shows with shift. The second and third rows show axons with 2D Gaussian smoothing perpendicular to the axon, and the latter two rows with 1D smothing longitudinally.

### Methods

Figure S1 gives an overview of the synthetic axons considered. First, a cylinder pointing along $\hat{\boldsymbol{z}}$ with length $L$ and radius $R_0$, where $L/R_0$ =15, was discretized on a 3D grid with $L_{grid} = 150 = 7.5R_0 = L/2$ grid points along each dimension. For each cross-sectional layer of the tube, the radius was perturbed according to a normal distribution with zero mean and standard deviation $R_0/4$. The synthetic axon was then hollowed by eroding each cross-sectional layer $R_0/3$ grid points and subtracting it from the original layer. Each layer was then randomly shifted $L_{shift}$ away from its center-of-mass (COM), again picked from a normal distribution with zero mean, while the standard deviation of the shift was varied across 7 individual simulations ranging from $L_{shift} = 0$ to $R_0/3$. To vary the structural correlation of the synthetic axon's surface morphology, we smoothed the axon by applying either a 1D, 2D or 3D Gaussian filter to the surface, with standard deviation ranging from $L_\sigma = 0$ to $R_0$ across 20 individual simulations. After smoothing, the synthetic axon was again represented by an indicator function $v_{axon}(\boldsymbol{r})$ (1 inside the layer and otherwise 0) by thresholding. The 2D filter was applied in the *xy*-plane perpendicular to the direction of the axon, while 1D filtering was done along the *z*-axis of the axon.

For each feature combination, we computed the induced Larmor frequency shift $\Omega(\boldsymbol{r})$ using Eq. (1) from the synthetic axon sheath described by the indicator function $v_{axon}(\boldsymbol{r})$. We then computed the magnetic field variance $\sigma^2$ inside and outside the synthetic axon $j = a, e$. $\Omega(\boldsymbol{r})$ was computed numerically, with zero padding perpendicular to the tube to avoid external fields leaking inside the tube and edge fields on the top and bottom of the synthetic axon. Hence, it is by zero-padding along the axial direction of the synthetic axon that its field appears to come from a tube twice the length of the 3D grid. The magnetizing external field $\boldsymbol{B}_0 = B_0\hat{\mathbf{B}}^{\mathrm{T}}$ was oriented along 100 unique orientations $\hat{\mathbf{B}}$ generated using electrostatic repulsion.

We also tested the effect of magnetized spherical inclusions inside the synthetic axon. Here we packed the intra-axonal space with randomly packed spheres all with a radius $R_S = R_0/4$. The density of spheres inside was kept at 10%. As for the tube, we numerically calculated $\Omega(\boldsymbol{r})$ from the spheres and computed the variance outside the sphere, and inside the synthetic axon, for each measurement direction $\hat{\mathbf{B}}$.

**Results**

Figure S2 shows the Larmor frequency variance $\sigma^2$ inside and outside the synthetic axons, respectively, for the different shifts and filters, while Figure S4 shows the fitting parameters. Here we found Eq. (9) could describe all cases considered.

Figure S3 shows $\sigma^2$ caused by the intra-axonal spheres while Figure S4 shows the fitting parameters. Here we see that, for all cases considered here, intra-axonal $\sigma^2$, appears dipolar behaving as $(1 - 3\cos^2(\theta))^2$. Outside the axons, which is isolated from the spheres by the synthetic myelin layer, the variance $\sigma^2$ goes as $\sin^4(\theta)$, which is sensible since the field from a cylindrical arrangement of spheres appear as that of a straight cylinder sufficiently far away from its axis. Hence, Eq. (9) can explain the magnetic field variance induced both inside and outside the synthetic axons, and from the spherical sources and myelin-like sheaths.

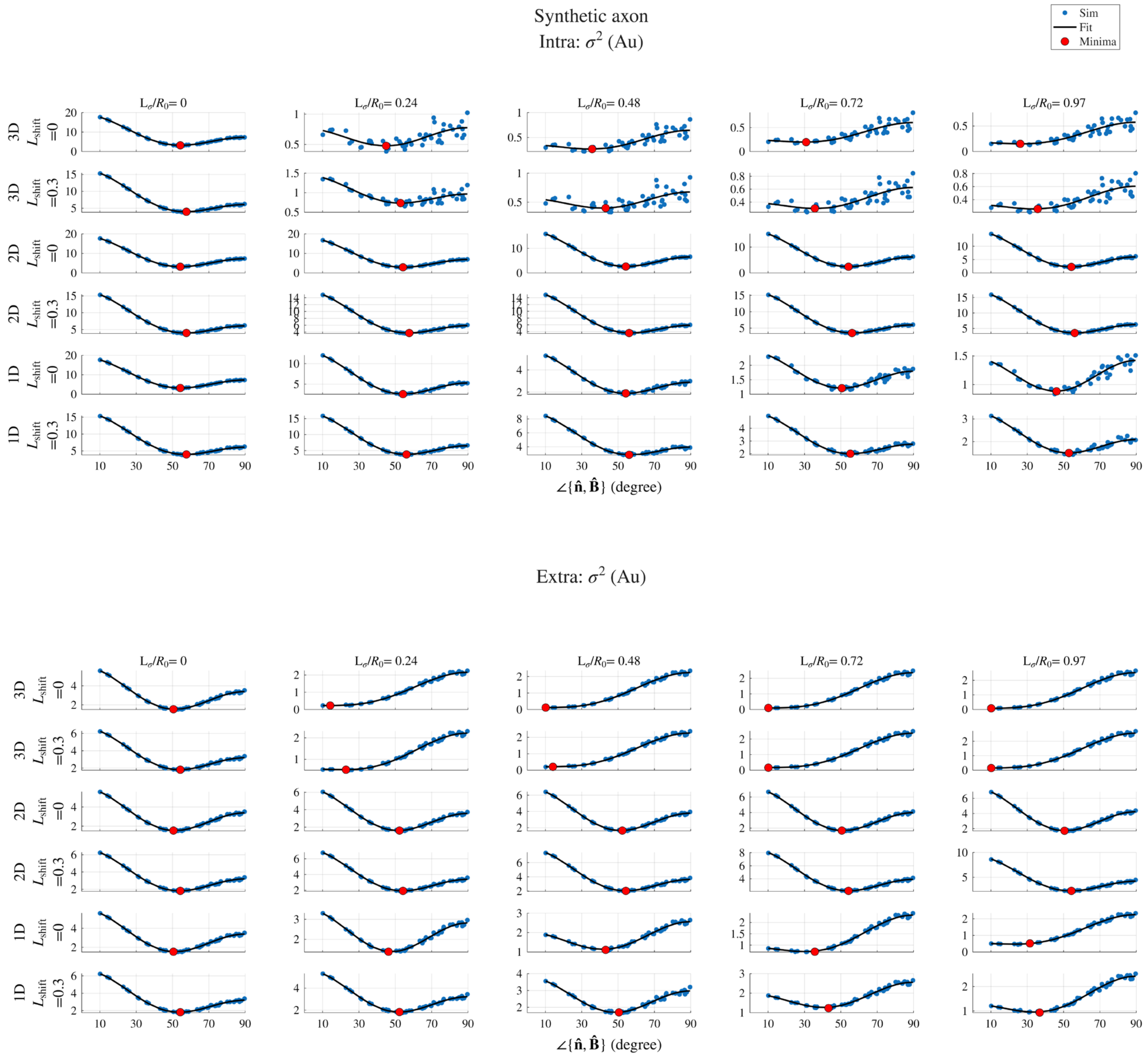

Figure S2 - Larmor frequency variance indcued by synthetic axonal myelin sheath with scalar susceptibility. First six rows show the intra-axonal magnetic field variance for 3D, 2D or 1D Gaussian smoothing and cross-sectional shift $L_{shift}$. The last six rows show variance outside the synthetic axons. X-axis denotes the angle between B0 and axon.

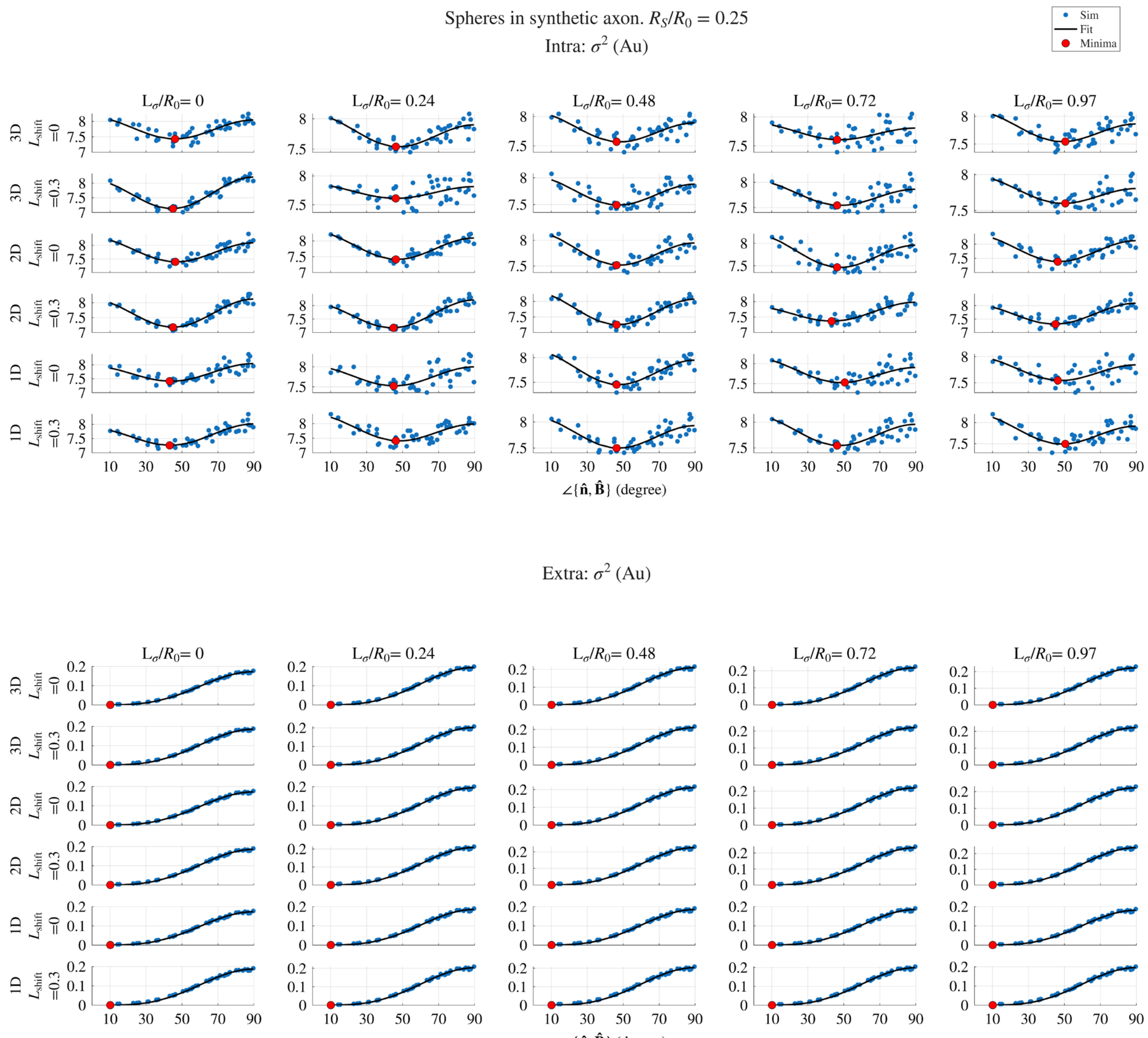

Figure S3 - Larmor frequency variance induced by randomly packed spheres with scalar susceptibility inside a synthetic axon. First six rows show the intra-axonal variance for 3D, 2D or 1D Gaussian smoothing and cross-sectional shift $L_{shift}$. The last six rows show variance outside the synthetic axons. X-axis denotes the angle between B0 and axon.

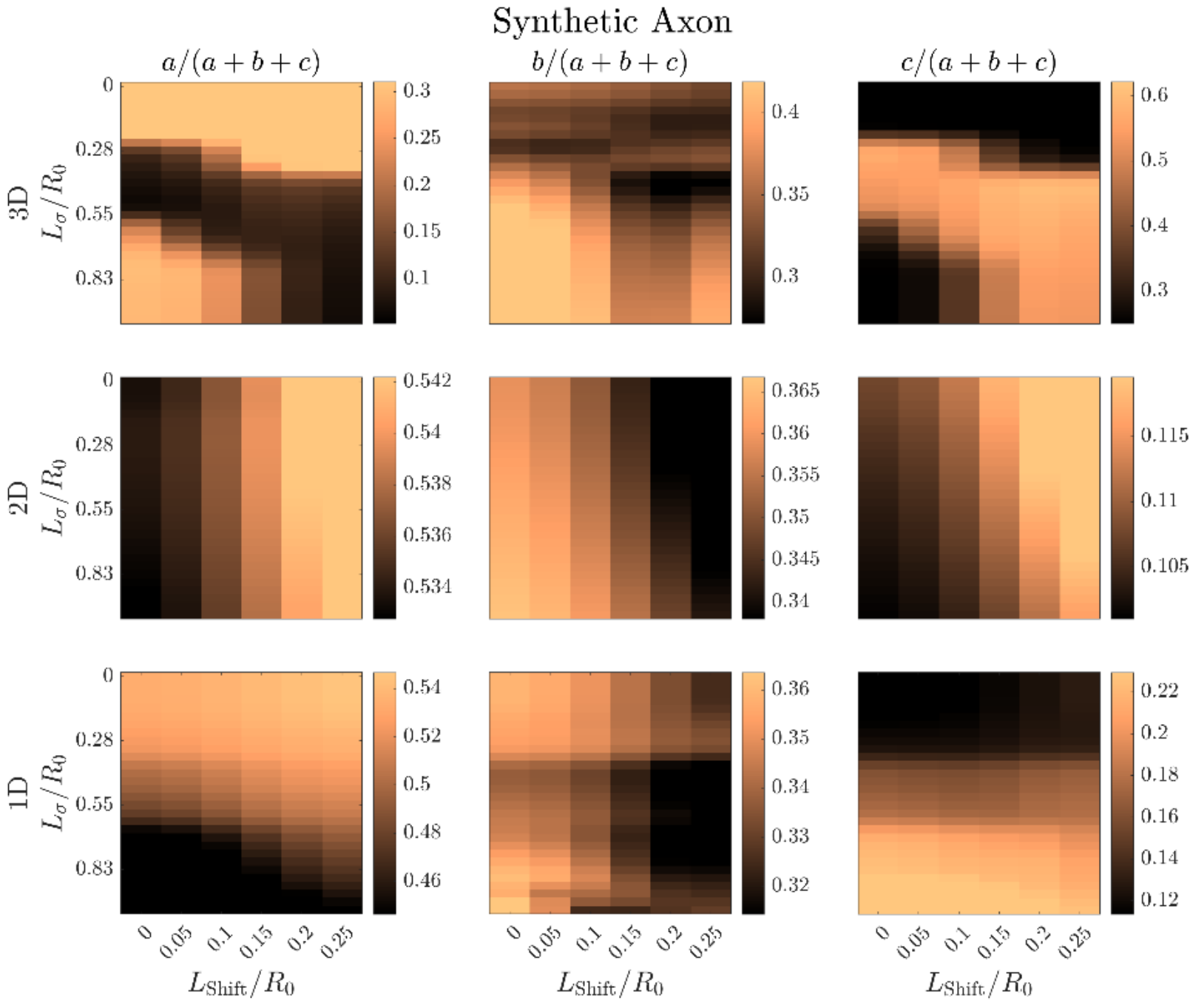

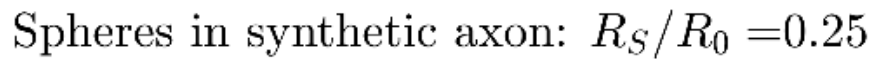

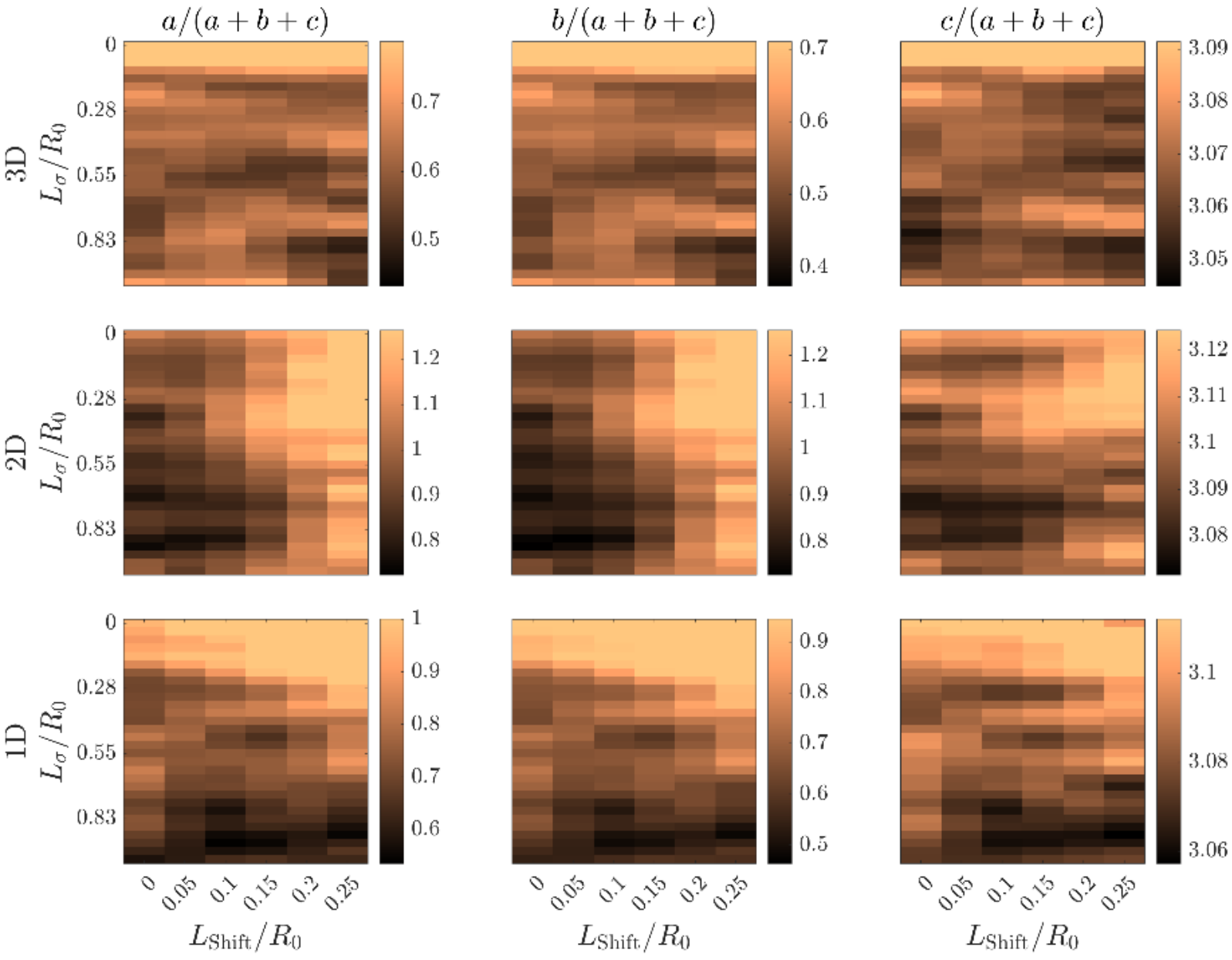

Figure S4 - Parameters of Eq. (9) after fitting Larmor frequency variance induced inside synthetic axon generated by either the synthetic axons or by randomly packed sphered in the axons. Here is shown for a radius $R_S/R_0 = 0.25$ for the spheres compared to the mean axon radius. X-axis denoted the amount of cross-section shifts induced for each axon slice, while the y-xais the size of the smoothing filter. First two rows show for 3D Gaussian smoothing, the next two 2D Gaussian smoothing and the latter rows 1D Gaussian smoothing. Colors are clippped in order to visualize the whole range of values.

# S2 - Larmor frequency variance inside hollow synthetic axons packed with spheres

We also designed a set of simulations to examine the orientation dependence of Eqs. (4) and (8) from *a*) individual WM axons segmented from electron microscopy (EM), and *b*) full WM axonal microstructures from EM containing thousands of axons. In both simulations, we considered the individual contribution from Larmor frequency shifts induced *i*) by the axons and *ii*) manually introduced intra-axonal spherical inclusions. Every inclusion, also myelin, was assumed to have scalar susceptibility(Sandgaard et al., 2024; Wharton & Bowtell, 2015) in every simulation. We denote the four cases as *ai), aii), bi)* and *bii),* respectively.

**Methods**

*a) Larmor frequency variance inside individual realistic axons with varying morphology*

In the spirit of Winther et al. (Winther et al., 2024) and Lee et al. (Lee et al., 2020), we considered the intra-axonal Larmor frequency variance variances $\overline{\sigma^2}$, $\sqrt{\overline{(\sigma^2)^2} - (\sigma^2)^2}$ and $\overline{\Omega^2} - \overline{\Omega}^2$, and extra-axonal variance $\overline{\sigma^2}$ for $B_0 = 7\text{T}$ and various orientations $\hat{\mathbf{B}}$. The frequency shift was induced by realistic axons with varying morphological features to assess how these structural variations influence the field variance and whether the resulting orientation dependence can be explained by Eq. (8). Figure 1 provides an overview of the axons analyzed, which were extracted from 8 different white matter substrates segmented from openly available electron microscopy (EM) data (see previous work(Sandgaard & Jespersen, 2025) for details). Two of the substrates were from tissue affected by traumatic brain injury (TBI). For each substrate, we identified the principal fiber direction and selected a major axon bundle consisting of approximately 2,000 axons aligned predominantly along this direction. As described in our previous study(Sandgaard & Jespersen, 2025), the axonal microstructure in each substrate was defined by an indicator function $v(\boldsymbol{r})$ on a 3D grid with resolution of 0.1 $\mu m^3$ (see ref (Sandgaard & Jespersen, 2025) for more details). The labels C2-C5 in Figure 1 are adopted from Winther et al.(Winther et al., 2024), while C6 is an additional case designed to isolate the effect of non-circular cross section. For that, we modified the center of mass for each cross-sectional slice along the main direction of the axon to be equal to remove any slowly varying axial tortuosity, while retaining the original non-circular cross-section. The extra-axonal space outside one axon was here defined by dilating the mask $v(\boldsymbol{r})$ such that the axon diameter was twice as large and then subtracting $v(\boldsymbol{r})$. For the myelinated axons in *ai)*, we used an intrinsic scalar susceptibility of $\chi_\text{m} = -100\frac{\text{ppb}}{\zeta_\text{m}}$ (Luo et al., 2014; Sandgaard et al., 2024; Wharton & Bowtell, 2015), where $\zeta_m$ is the volume fraction of axons in the entire microstructure. The local Larmor frequency shift $\Omega(\boldsymbol{r})$ was calculated numerically(Ruh & Kiselev, 2018) using Eq. (1), with $\chi_m(\boldsymbol{r}) = \chi_m v_m(\boldsymbol{r})$ for 50 unique orientations of $\hat{\mathbf{B}}$ generated using electrostatic repulsion(Jones et al., 1999). As our simulation resolution (0.1 μm³) was too coarse to model effects from small inclusions like ferritin molecules with a radius around 4 nm,(Duyn & Schenck, 2017) we instead modelled iron-containing cells in *aii)* with a radius of approximately 0.5 μm. We chose a volume fraction of $\zeta_s = 0.05$, and an intrinsic spherical susceptibility of $\chi_s \approx 1100$ ppb, such that the bulk susceptibility in the whole microstructure was $\overline{\chi}_s = \zeta_s \chi_s \approx 55$ ppb. The susceptibility, size and volume fraction could mimic dopaminergic cells containing e.g. ferritin and neuromelanin(Brammerloh et al., 2021), and may be found in WM near the Substantia Nigra(Brammerloh et al., 2021). Non-overlapping intra-axonal spheres were randomly packed using a previously designed packing generator(Sandgaard et al., 2023). We assumed the spherical cellular inclusions were impenetrable to water and their water signals fully relaxed, like myelin water. Again, we calculated the local Larmor frequency shift $\Omega(\boldsymbol{r})$ numerically(Ruh & Kiselev, 2018) using Eq. (1), with $\chi_s(\boldsymbol{r}) = \chi_s v_s(\boldsymbol{r})$ for the same 50 orientations of $\hat{\mathbf{B}}$.

b) *Larmor frequency variance inside axonal bundle of realistic axons*

While the previous simulation considered the self-induced frequency variance from every individual axon only, we next considered $\overline{\sigma^2}$, $\sqrt{\overline{(\sigma^2)^2} - \left(\overline{\sigma^2}\right)^2}$ and $\overline{\Omega^2} - \overline{\Omega}^2$ in the intra-axonal space, and $\overline{\sigma^2}$ in the extra-axonal space, caused by the entire substrate for all 50 orientations of $\hat{\mathbf{B}}$. This was done *bi)* for the axonal microstructure, and *bii)* for intra-axonal spherical inclusions. Hence, the difference between simulation *b)* compared to *a)* is that here, induced fields from neighboring axons or spheres are present. To compute the extra-axonal field variance in *b*), we dilated the intra-axonal mask, similar to simulation *a)*, but now of the ~2000 axons in the bundle. We then multiplied the dilated intra-axonal mask with the negated original masks of myelin and intra-axonal space to segment the extra-axonal space in the vicinity of the axonal bundle. We reused the sphere packing from *a)*, but here we computed the induced field from all the spheres inside the intra-axonal space in the entire microstructure shown in Figure S5.

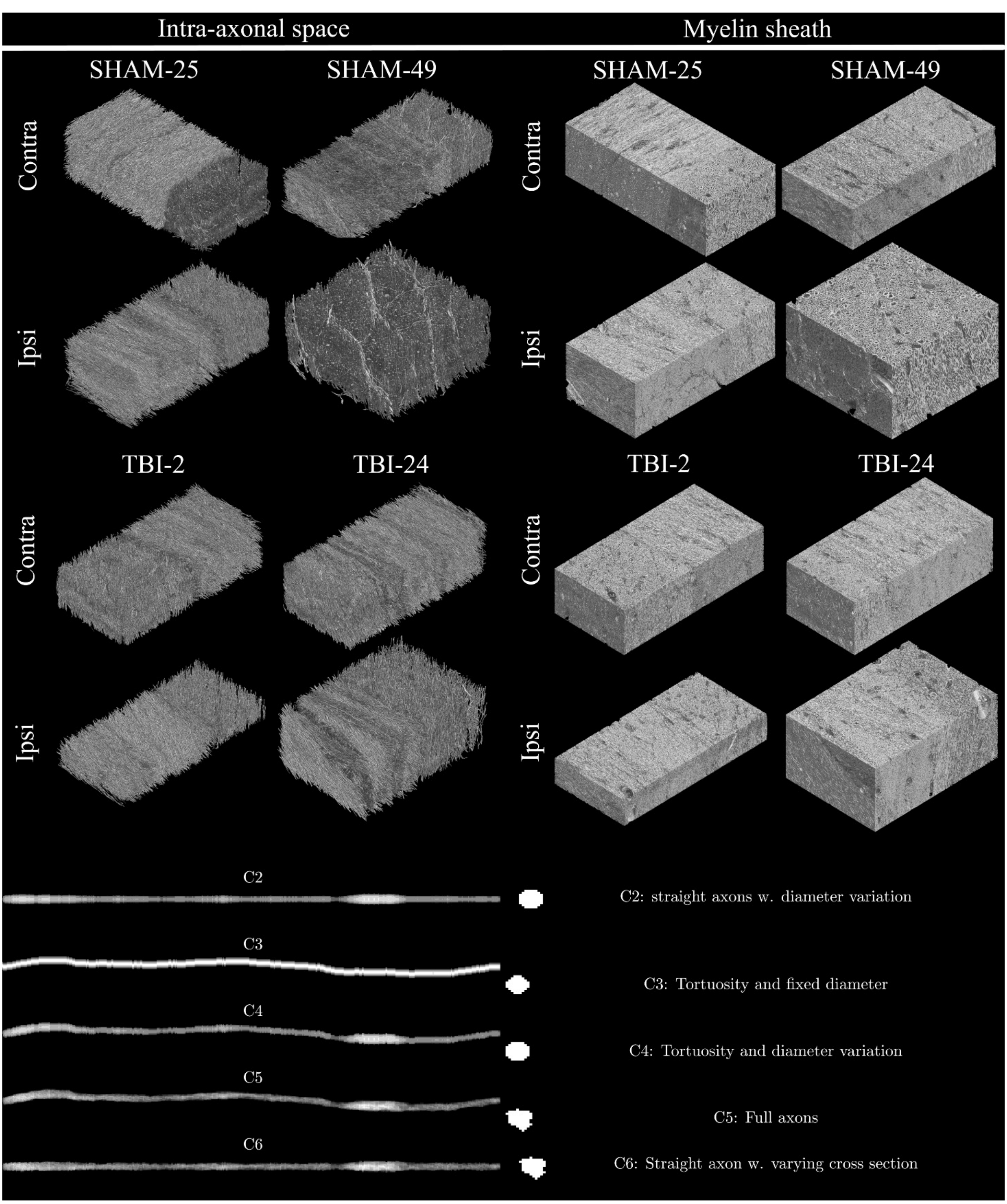

Figure S5 - In-silico white matter axon phantoms used for Monte-Carlo simulations. Eight different substrates from two different SHAM rats labelled 25 and 49 and two different TBI rats labelled 2 and 24 were used for Monte-Carlo simulations. Labels correspond to the original data. For each brain, both ipsilateral (ipsi) and contralateral (contra) tissue samples are considered. The tissue is extracted from the corpus callosum and cingulum bundles. The intra-axonal spaces are used for the Monte-Carlo simulation of diffusing spins, while the myelin sheaths constitute the magnetizable tissue, perturbing the Larmor frequency of the diffusing spins.

For each substrate, we considered the magnetic field variance induced by each axon. In the spirit of Winther et

al.(Winther et al., 2024) we synthesized 6 different axons with varying microstructural features to investigate the intra-axonal and extra-axonal magnetic field average and variance. Labeling was kept as in Winther et al. for consistency.

## Results

*a) Magnetic field variance inside realistic axons with varying morphology*

Figure S6 shows $\overline{\sigma^2}$, $\sqrt{\overline{(\sigma^2)^2} - \left(\overline{\sigma^2}\right)^2}$ and $\overline{\Omega^2} - \overline{\Omega}^2$ for realistic axons with varying morphology (C2-C6). Eq. (8) consistently captures the orientation dependence across all cases, with parameter values reflecting differences in axonal morphology. Remarkably, none of the simplistic axons are close to the orientation dependence observed in the full axons (C5), which means that all structural features are important when modeling relaxation in axons. Outside the axons, the angular dependence of the magnetic field variance induced by intra-axonal spheres is similar to that of a long cylinder ($\sin^4(\theta) = 1 - 2\cos^2(\theta) + \cos^4(\theta)$).

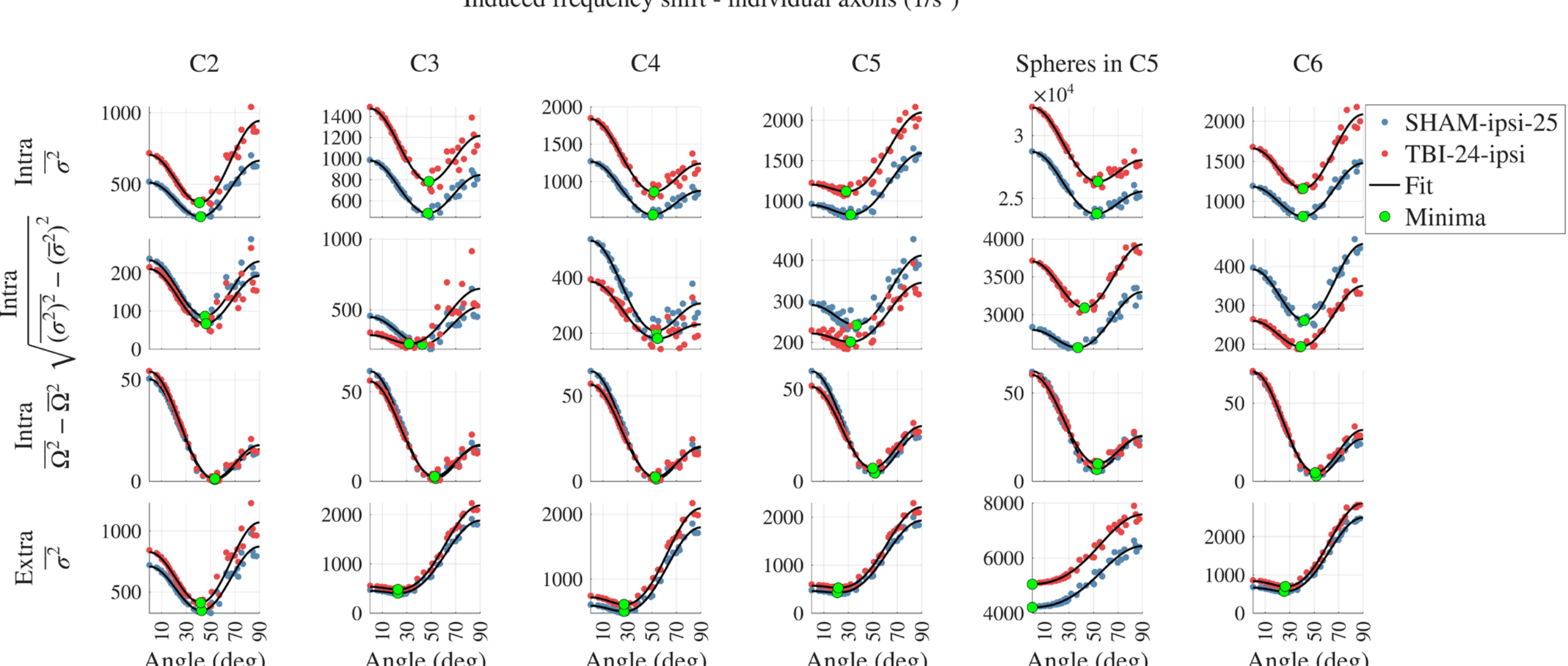

Figure S6– Larmor frequency variance induced by individual axonal myelin sheaths or intra-axonal spherical sources (C5 spheres) versus the angle between B0 and average direction of the axon. First three rows show the variance inside every axon, while the last row shows extra-axonal variance. Labels (C2-C6) indicate the diffent morphological features considered (see Figure 1). Each line shows results for 2 different WM substrates, including fits and minima.

*b) Magnetic field variance inside microstructure of realistic axons with varying morphology*

Figure S7 shows the results for the fields generated by the full microstructure of 2 substrates, in terms of variances within and outside the major fiber bundle. Interestingly, we find upon comparison with Figure S6 that the field variances $\overline{\sigma^2}$, $\sqrt{\overline{(\sigma^2)^2} - \left(\overline{\sigma^2}\right)^2}$ and $\overline{\Omega^2} - \overline{\Omega}^2$, acquires a non-negligible contribution from the other magnetized axons. This makes intra-axonal $\overline{\sigma^2}$ more cylinder-like as it behaves as $\sin^4(\theta)$.

The variance $\overline{\sigma^2}$ in Figure S7 caused by intra-axonal spheres from the whole microstructure is comparable in magnitude with the self-induced variance $\overline{\sigma^2}$, as seen in Figure S6. This means that the sphere-induced frequency shift variance largely comes from sources inside the axons. This is also clear upon looking at the extra-axonal variance going as $\sin^4(\theta)$, which is around 5 times weaker in magnitude compared to the variance induced in the intra-axonal space. Equation (8) could fit all cases well.

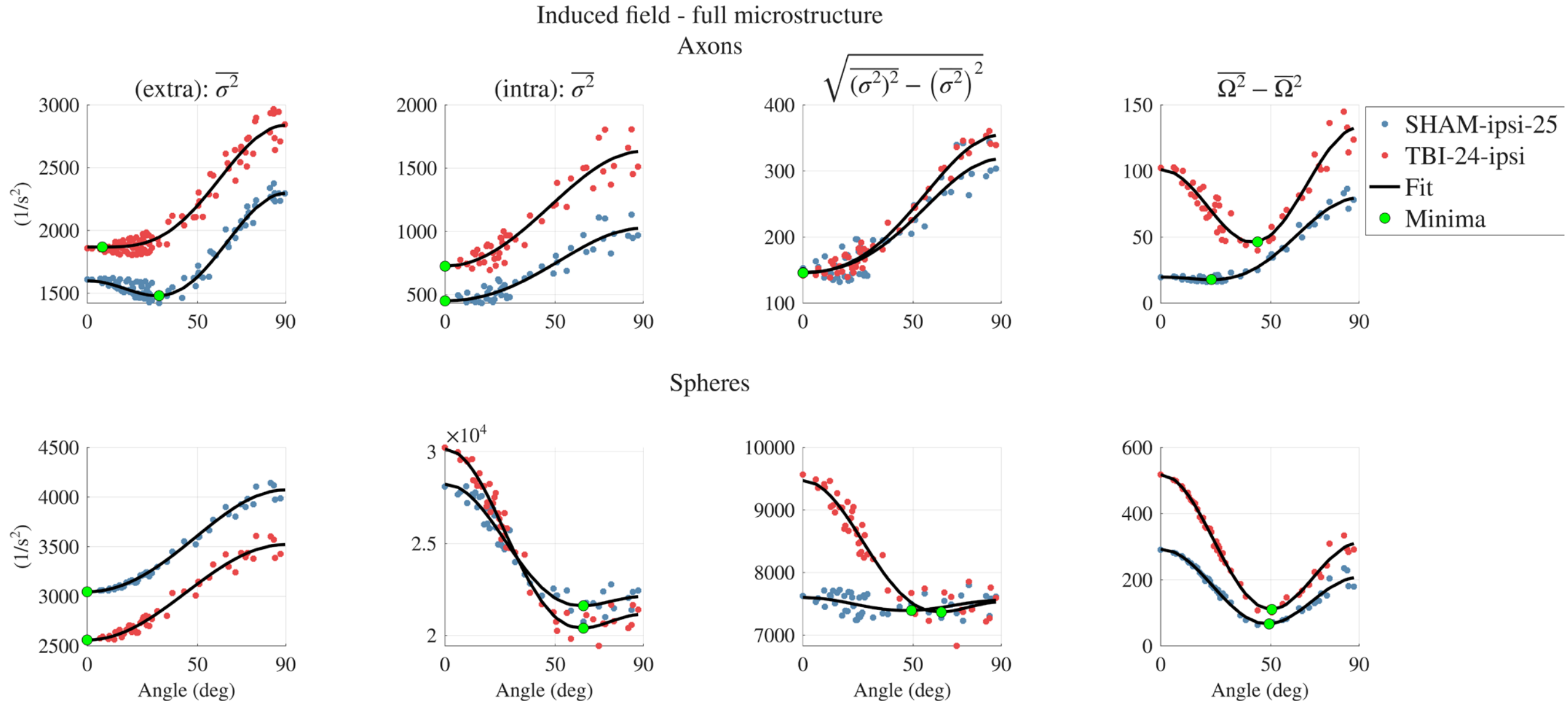

Figure S7 - Larmor frequency variance, induced by axonal myelin sheaths with scalar susceptibility, versus angle between B0 and the average direction of the major fiber bundle. First row shows the intra-axonal variance for 2 different WM substrates, with the first column displays the variance in the extra-axonal space. Second row shows the variance by intra-axonal spherical sources with scalar susceptibility.

# S3 - MR Signal from multiple mesoscopic regions

Relating the total signal decay function to a second order cumulant expansion across water compartments with different decay functions $\eta_c$ is justified only when the variability in $\eta_c$ and $\Omega_c$ across compartments is sufficiently small. However, in a macroscopic voxel, e.g. in WM, there exist multiple bundles of axons oriented differently to the external $\mathbf{B_0}$ field, which can challenge the validity of this compartmental expansion. Second, water in the extra-axonal space in WM may also experience a different total decay function $\eta_e$, which requires separation of intra- and extra-axonal space on the mesoscopic scale. Hence, if we consider a macroscopic volume consisting of multiple mesoscopic sub-volumes $\mathcal{M}$, where the mesoscopic signal $\mathcal{K}_{\mathcal{M}}$ is well approximated by the second order expansion across its compartments, e.g. coherent bundles of axons with different orientations $\hat{\boldsymbol{n}}$, the net signal becomes a sum

$$S = \sum_{\mathcal{M}} \mathcal{K}_{\mathcal{M}} = \sum_{\mathcal{M}} \exp(-\eta_{\mathcal{M}} - i\varphi_{\mathcal{M}}). \tag{1}$$

(multiple mesoscopic regions)

Here the notation $\mathcal{K}_{\mathcal{M}}$ is motivated by the Standard Model of diffusion (SM) in WM(Novikov et al., 2019). Applying this to WM, we consider the axonal microstructure as a random medium consisting of many orientationally dispersed bundles of aligned axons. We categorize the MR fluids as intra-axonal ($a$) and extra-axonal water ($e$), where every axon bundle is assumed to be described by the same signal kernel (myelin water is assumed to be fully relaxed). As axonal bundles can have different orientations $\hat{\boldsymbol{n}}$, we replace the sum over $\mathcal{M}$ with an integration over a fiber orientation distribution function $\mathcal{P}(\hat{\boldsymbol{n}})$,

$$S = S_0 \int d\hat{\boldsymbol{n}}\, \mathcal{P}(\hat{\boldsymbol{n}}) \mathcal{K}(\hat{\boldsymbol{n}}), \tag{2}$$

where

$$\mathcal{K}(\hat{\boldsymbol{n}}) = f_a \exp\big(-\eta_a(\hat{\boldsymbol{n}}) - i\varphi_a(\hat{\boldsymbol{n}})\big) + (1 - f_a) \exp\big(-\eta_e(\hat{\boldsymbol{n}}) - i\varphi_e(\hat{\boldsymbol{n}})\big) \tag{3}$$

(WM signal)

is the mesoscopic signal kernel and depends on the angle between the axons and external field. The decay functions $\eta_a$ and $\eta_e$ is described by Eqs. (4) and (8), but $\eta_e$ has no contribution from inter-compartmental variance, since the extra-axonal space only consists of one water compartment, and hence all variance is intra-compartmental per definition. As we assume each bundle is characterized by the same signal kernel $\mathcal{K}$, the transverse relaxation rate in each bundle originates *statistically* from similar local microstructure, only oriented differently with respect to the external field. The transverse relaxation induced between neighboring bundles is also orientation dependent, but contributes equally to each bundle, as the microstructure external to each bundle should look the same assuming a translation invariant random medium. Transverse relaxation from other field variations also affects the signal - both on the molecular scale (e.g., dipole-dipole relaxation(Bloembergen et al., 1948; Yablonskiy & Sukstanskii, 2025)) and macroscopic scale (e.g., macroscopic variations across the extent of the point-spread-function(Yablonskiy et al., 2013)), but these are not considered in this study.

# S4 - Supplementary figures

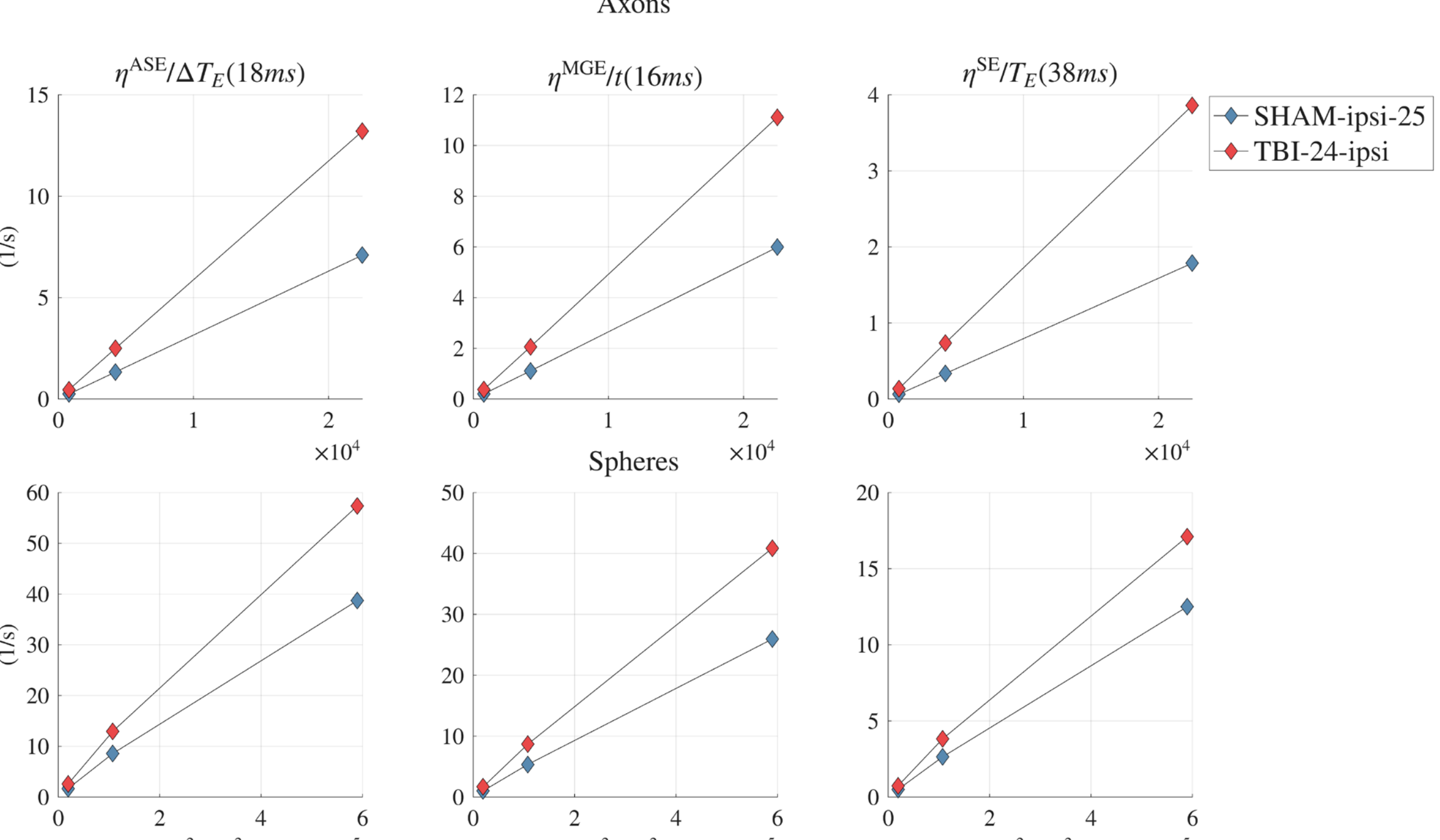

Figure S8 - Transverse relaxation for spin-echo (SE), asymmetric spin-echo (ASE) and multi-gradient echo (MGE) signals plotted against the $B_0$ strength squared, and angled at 65 degrees to the axon bundle. Colored points correspond to 2 different axonal substrates. The first row shows the relaxation induced by the realistic axonal microstructure. Bottom row shows the relaxation induced by spheres.

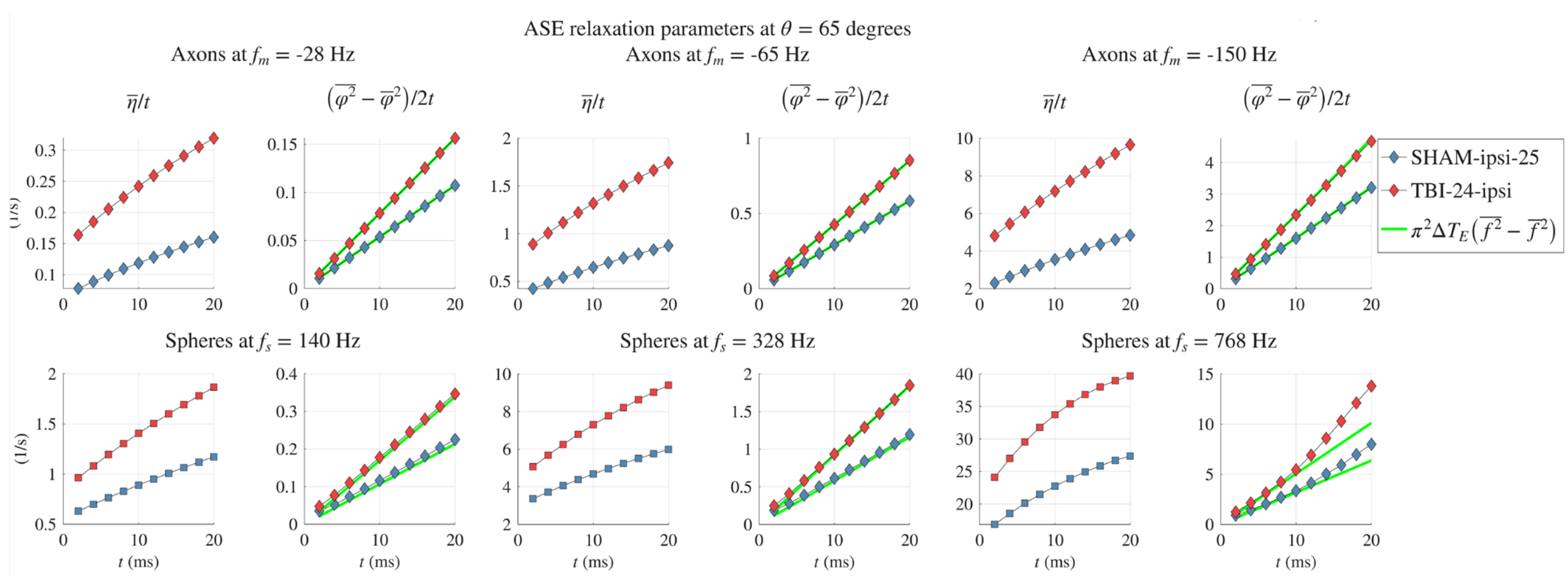

Figure S9 - Fitting parameters from fitting the ASE signal decay to the magnetic field variances for different echo times $\Delta T_E$ (Eq. (4)). Colors correspond to 2 different axonal substrates. The first row shows the contribution induced by the full EM axonal microstructure, while the bottom for spheres. Relaxation is shown for one characteristic frequency. The green lines in the third column show the estimated phase variance from the induced magnetic field variance $\Delta \boldsymbol{B}$.

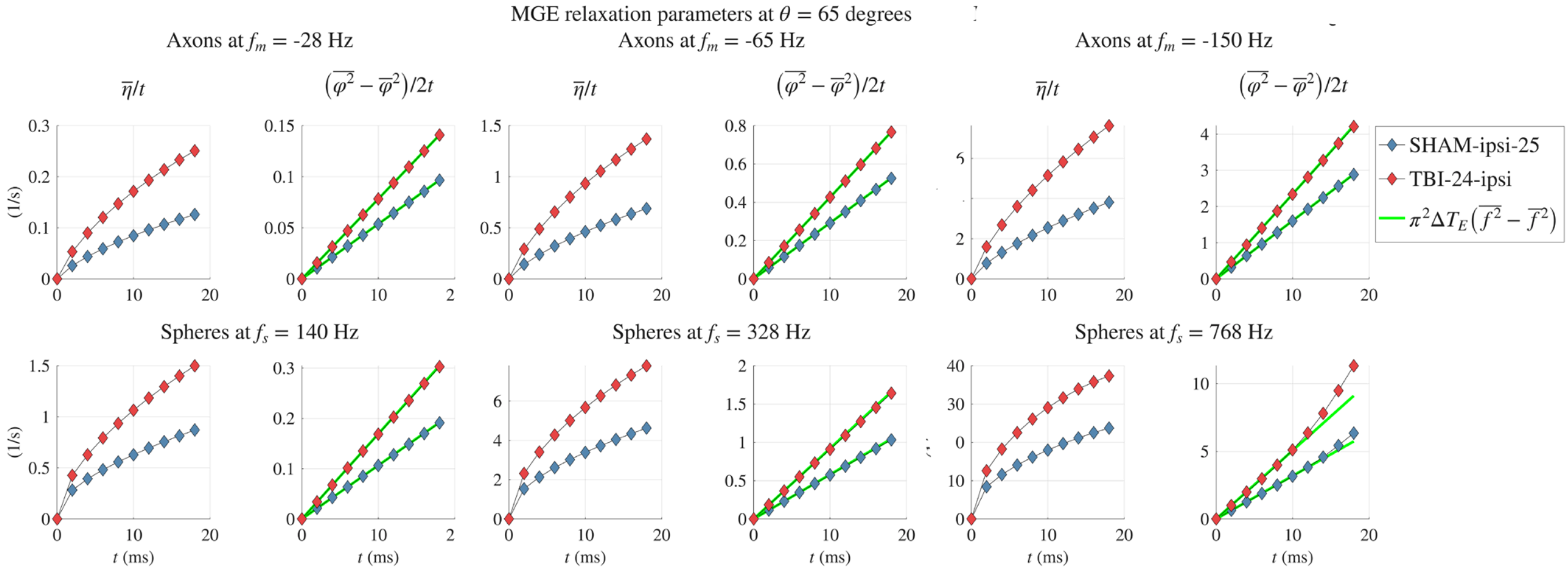

Figure S10 - Fitting parameters from fitting the MGE signal decay to the magnetic field variances for different echo times $t$, (Eq. (4)). Colors correspond to 2 different axonal substrates. The first row shows the contribution induced by the axonal microstructure, while the bottom is for sphere-filled axons. Relaxation is shown for one characteristic frequency. The green lines in the third column show the estimated phase variance from the induced magnetic field variance $\Delta \boldsymbol{B}$.

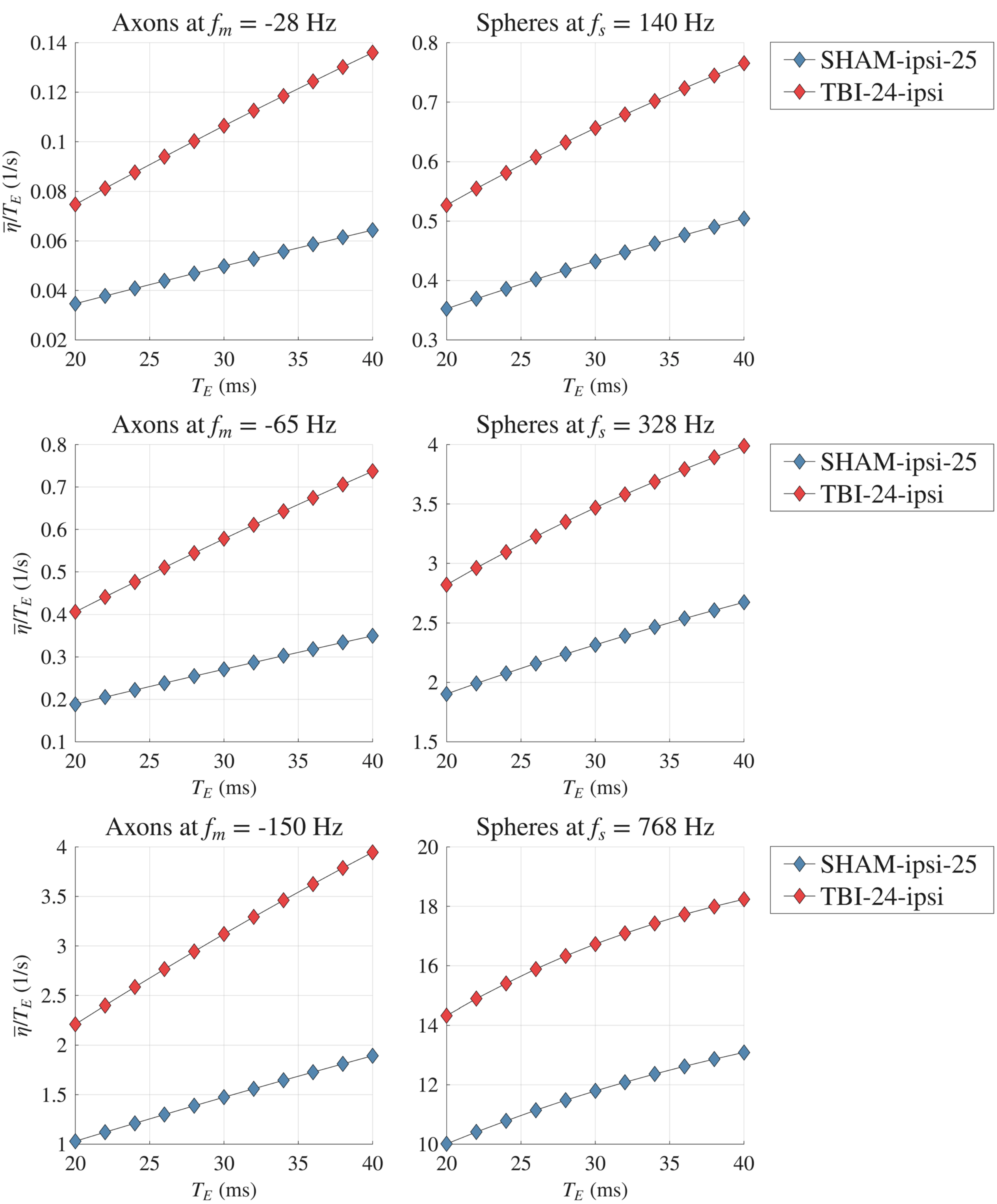

Figure S11 - Fitting parameters from fitting the SE signal decay to the magnetic field variances for different echo times $T_E$ and $\Delta T_E = 0$, (Eq. (5)). Colors correspond to 2 different axonal substrates. The first row shows the contribution

induced by the axonal microstructure, while the bottom is for sphere-filled axons. Relaxation is shown for one characteristic frequency.

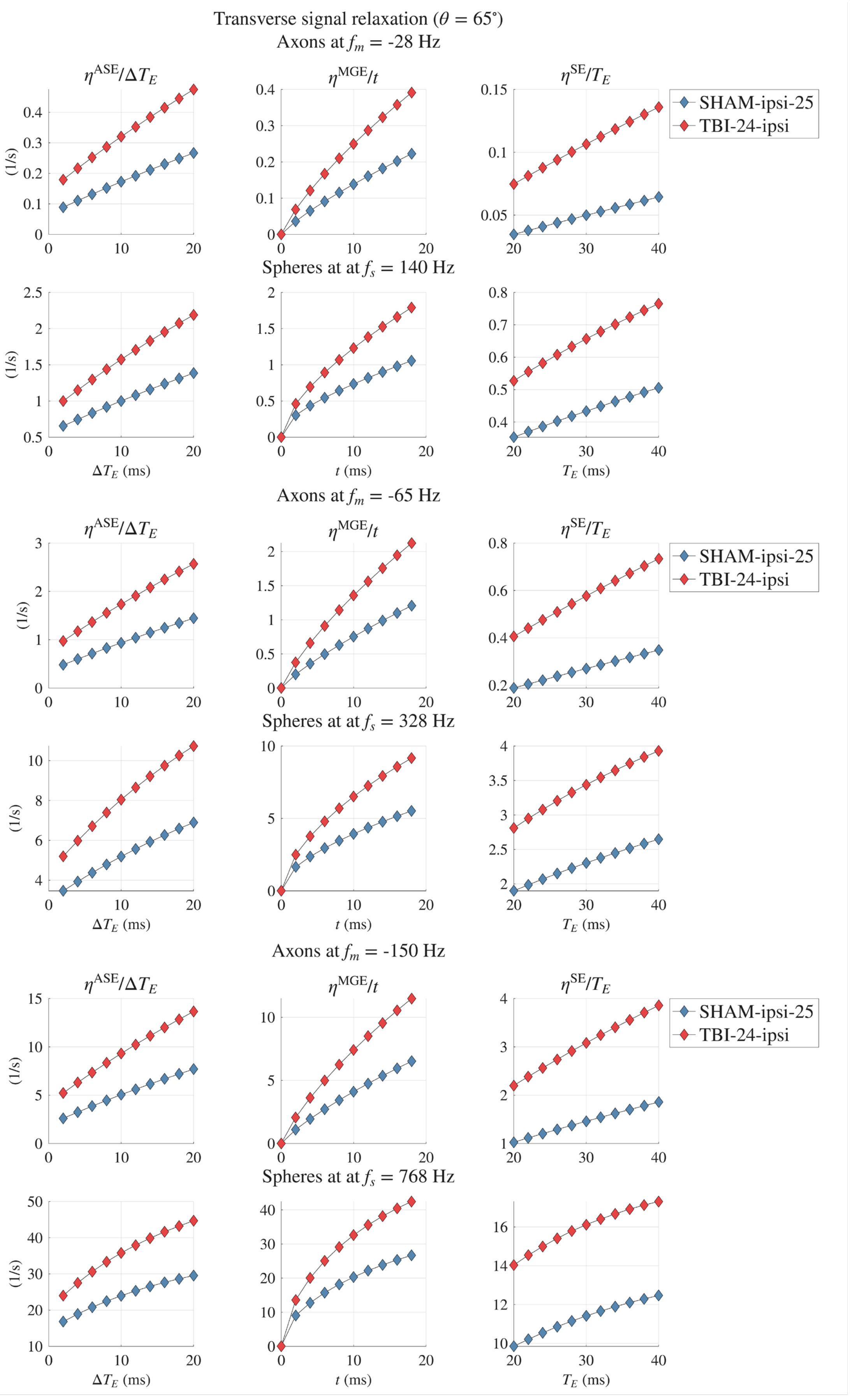

Figure S12 – Transverse relaxation for an asymmetric spin-echo (ASE) and multi-gradient echo (MGE) signal plotted against echo time. Colors correspond to 2 different axonal substrates. The external field is oriented 65 degrees to the main fiber direction. The first row shows the signal relaxation induced by the unmodified EM axonal microstructure. Relaxation is shown for one characteristic frequency. Bottom row shows the relaxation induced by spheres.

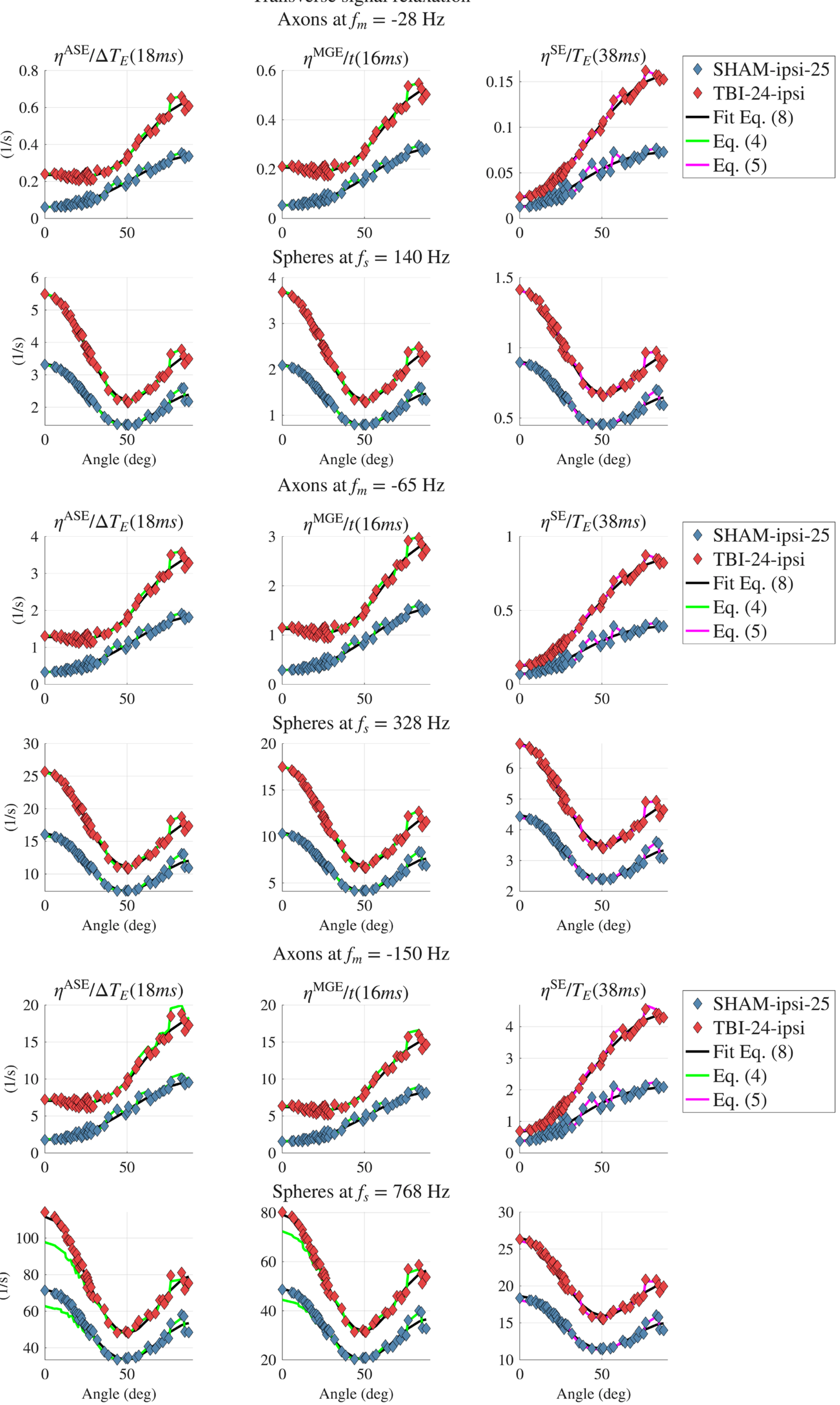

Figure S13 - Transverse relaxation for spin-echo (SE), asymmetric spin-echo (ASE) and multi-gradient echo (MGE) signals plotted against the angle between the external field and the main fiber bundle at a fixed echo time. Relaxation is shown for one characteristic frequency. Colored points correspond to 2 different axonal substrates. The first row shows the relaxation induced by the realistic axonal microstructure. Bottom row shows the relaxation induced by spheres. Black line shows fitting to Eq. (8) and the green line estimated from Eq. (4) for MGE and ASE and Eq. (5) for SE in magenta.